\pdfoutput=1
\documentclass[a4paper,12pt]{article}
\usepackage{amsfonts}
\usepackage{mathrsfs}
\usepackage{amsmath}
\usepackage{amssymb}
\usepackage[medium]{titlesec}
\usepackage{bm}
\usepackage{cite}
\usepackage[normalem]{ulem}
\usepackage{extarrows}
\usepackage{slashed}
\usepackage{isodateo}
\usepackage{graphicx}
\usepackage{xcolor}
\usepackage[bookmarksnumbered=true,bookmarksopen=true]{hyperref}
 \hypersetup{colorlinks,%
             linkcolor=[rgb]{0,0.2,0.6}, %
             citecolor=[rgb]{0,0.2,0.6}}
\usepackage[hmargin=.7in,vmargin=1.1in]{geometry}
\usepackage{indentfirst}
\newcommand{\FR}[2]{\displaystyle\frac{\,{#1}\,}{#2}}

\newcommand{\n}{\nonumber}

\graphicspath{{fig/}}

\def\bge{\begin{equation}}
\def\ede{\end{equation}}
\def\bga{\begin{aligned}}
\def\eda{\end{aligned}}
\def\bgp{\begin{pmatrix}}
\def\edp{\end{pmatrix}}
\def\bgs{\begin{subequations}}
\def\eds{\end{subequations}}
\newcommand{\order}[1]{\mathcal{O}({#1})}
\def\di{{\mathrm{d}}}

\def\mb{\mathbf}

\def\pd{\partial}
\def\ld{{\mathscr{L}}}

\def\la{\langle}\def\ra{\rangle}

\def\to{\rightarrow}

\def\ii{\mathrm{i}}

\def\de{\delta}

\def\ka{\kappa}
\def\lam{\lambda}

\def\si{\sigma}

\def\Re{\mathrm{Re}\,}
\def\Im{\mathrm{Im}\,}

\newcommand{\wt}[1]{\mkern 2mu \widetilde{\mkern -2mu #1 \mkern -2mu}\mkern 2mu}

\begin{document}

\title{ \Large\textbf{Quantum Standard Clocks in the Primordial Trispectrum}}
\author{
Xingang Chen$^{a}$,~
Wan Zhen Chua$^{b}$,~
Yuxun Guo$^{c}$,\\
Yi Wang$^{b,d}$,~
Zhong-Zhi Xianyu$^{e,f}$,~
Tianyou Xie$^{c}$
\\[3mm]
\normalsize{$^a$~\emph{Institute for Theory and Computation, Harvard-Smithsonian Center for Astrophysics,}}\\
\normalsize{\emph{60 Garden Street, Cambridge, MA 02138, USA}}\\
\normalsize{$^b$~\emph{Department of Physics, The Hong Kong University of Science and Technology,}}\\
\normalsize{\emph{Clear Water Bay, Kowloon, Hong Kong, P.R.China}}\\
\normalsize{$^c$~\emph{Department of Physics, Tsinghua University, Haidian DS, Beijing 100084, China}}\\
\normalsize{$^d$~\emph{Jockey Club Institute for Advanced Study, The Hong Kong University of Science and Technology,}}\\
\normalsize{\emph{Clear Water Bay, Kowloon, Hong Kong, P.R.China}}\\
\normalsize{$^e$~\emph{Department of Physics, Harvard University, 17 Oxford St., Cambridge, MA 02138, USA}}\\
\normalsize{$^f$~\emph{Center of Mathematical Sciences and Applications, Harvard University,}} \\
\normalsize{\emph{20 Garden St., Cambridge, MA 02138, USA}}\\
}
\date{}
\maketitle

\vspace{0.2cm}
\begin{abstract}

We calculate the primordial trispectrum of curvature perturbation in quasi-single field inflation, with general sound speeds for both the inflaton and the massive scalar. Special attention is paid to various soft limits of the trispectrum, where the shape function shows characteristic oscillatory pattern (known as the quantum primordial standard clock signal) as a function of the momentum ratio. Our calculation is greatly simplified by using the ``mixed propagator'' developed under a diagrammatic representation of the in-in formalism.

\end{abstract}

\newpage

\section{Introduction}
\label{sec_intro}

Primordial fluctuations of the spacetime at the very early stage seeded the inhomogeneity and anisotropy of our universe at large scales. Cosmological observations revealed that scalar fluctuations follow a nearly scale invariant and nearly Gaussian distribution. On the other hand, some non-Gaussian components in the primordial fluctuations are generally expected to exist and could be detected in the future observations by measuring the correlation functions of primordial fluctuations \cite{Bartolo:2004if,Liguori:2010hx,Chen:2010xka,Wang:2013eqj}. The primordial non-Gaussianity encodes a wealth of information about the dynamics and interactions at high energy, and provides an exciting way of searching for new physics.

A particularly attractive scenario that can generate large non-Gaussianities is the quasi-single field inflation (QSFI)
\cite{Chen:2009we,Chen:2009zp,Baumann:2011nk,Chen:2012ge, Sefusatti:2012ye, Norena:2012yi, Noumi:2012vr, Gong:2013sma, Emami:2013lma, Kehagias:2015jha, Arkani-Hamed:2015bza,  Dimastrogiovanni:2015pla,Chen:2015lza, Chen:2016cbe,Lee:2016vti,Chen:2016qce, Meerburg:2016zdz,Chen:2016uwp,Chen:2016hrz,
An:2017hlx,An:2017rwo,
Iyer:2017qzw, Kumar:2017ecc, Franciolini:2017ktv, Tong:2018tqf,MoradinezhadDizgah:2018ssw, Saito:2018xge}, where one or more scalar fields with mass $m$ around the Hubble scale $H$ are present along with the inflaton. Unlike the highly suppressed self-interaction of the inflaton due to the slow-roll condition, the self-interactions among the massive scalars in QSFI are not suppressed a priori and can become large. The large self-interaction can in turn be translated to large non-Gaussianities of the inflaton via the coupling between the inflaton and the massive fields.

More interestingly, in certain soft limits of $n$-point functions of the inflaton ($n\geq 3$), the massive scalar field can generate a characteristic oscillation/scaling shape through its interaction with the inflaton \cite{Chen:2009we,Chen:2009zp}. Similar patterns also show up for massive fields with nonzero spins that couple to the inflaton \cite{Arkani-Hamed:2015bza,Lee:2016vti}. This pattern is of particular interest for at least two reasons.

First, it tells us the (dressed) mass and spin of the massive fields that couple to the inflaton. Given that the Hubble scale can be extremely high during inflation, this turns the non-Gaussianity a rather clean channel for studying particle physics at the inflation scale, a viewpoint dubbed as the ``Cosmological Collider'' \cite{Arkani-Hamed:2015bza, Lee:2016vti, Chen:2016nrs,Chen:2016uwp,Chen:2016hrz}.

On the other hand, as pointed out in \cite{Chen:2015lza}, these signals are induced by quantum oscillations of massive fields that are periodic as a function of physical time, serveing as a ``quantum primordial standard clock''.\footnote{See also the classical primordial standard clock \cite{Chen:2011zf,Chen:2011tu,Chen:2012ja, Chen:2014joa,Chen:2014cwa}, where classical oscillations of the massive field are used as the standard clock.} Observing the physical-time-clocked oscillations in a range of cosmic scales (which translates to their conformal time of interaction with the massive field) reveals the relation between physical time and conformal time in the primordial universe. Thus, the oscillations can be used to measure the expansion or contraction history of the primordial universe, and distinguish inflation from alternative-to-inflation scenarios. For this reason, we will also call this oscillation pattern the ``clock signal'' in this paper.

It is thus important to work out explicitly the clock signals in $n$-point correlation functions of QSFI. Though this is seemingly a straightforward exercise under the standard in-in formalism, it turns out to be more than tedious to carry out the integrals directly. Apart from the difficulty of the time-ordering in the propagators, the numerical evaluation of the in-in integrals is plagued with divergences in various integration limits. The calculation of the bispectrum in \cite{Chen:2017ryl} was greatly simplified by the path-integral based diagrammatic approach -- the result that was two pages using the traditional method is now reduced to two lines; and the speed for numerical calculation is reduced from hours to seconds. The simplifications enable us to proceed in studying features of the bispectrum more carefully, such as limits of higher-point correlation functions related to halo bias \cite{An:2017hlx, An:2017rwo}.

In this paper, we apply the method in \cite{Chen:2017ryl} to QSFI and calculate the trispectrum (4-point correlation function), paying special attention to the clock signals in various soft limits \cite{Assassi:2012zq}, and provide simple semi-analytical results for the clock signals so that they can be readily evaluated numerically. The trispectrum for general single field inflation \cite{Chen:2006nt} has been extensively studied \cite{Chen:2009bc, Arroja:2009pd}. The relation between the sizes of trispectrum and bispectrum is different in QSFI and general single field inflation. Fixing the size of bispectrum, the trispectrum from the scalar-exchange diagram of the weakly-couplied QSFI is much greater than that of general single field inflation.\footnote{For the contact diagram of the trispectrum where four fields interact at the same spatial position, general single field and QSFI have similar potential to produce large trispectra. The trispectra can become very large while keeping the size of bispectrum for the following reasons. The coupling of the 4-point vertex can be made large independently \cite{Chen:2009bc, Arroja:2009pd} because it does not contribute to the 3-point coupling through loop diagrams, or one can choose to impose an approximate parity symmetry on the interactions \cite{Senatore:2010jy}.} This is because the amplitude of the general single field trispectrum is approximately that of the bispectrum squared. For QSFI, it can be much larger \cite{Chen:2009zp,Barnaby:2011pe}.

For generality, we consider general sound speed for both the inflaton and the massive isocurvaton. Inflation with different sound speeds for multiple light fields has been studied in \cite{Cai:2009hw}. For QSFI, in addition to the motivations given in \cite{Cai:2009hw}, it is worth noticing that integrating out additional massive fields modifies the sound speeds of lighter fields \cite{Tolley:2009fg, Achucarro:2012sm}. Also, the coupling between the inflaton and the isocurvaton modifies sound speeds \cite{Cremonini:2010ua, Iyer:2017qzw}. Thus the inclusion of general sound speeds adds more flexibility to the study.

The paper is organized as follows: We review the QSFI, and the diagrammatic in-in formalism with mixed propagator, in Sec.\;\ref{sec_QC}, where we also calculate the power spectrum and the bispectrum. We present the results of trispectrum and the clock signals in two soft configurations in Sec.\;\ref{sec_Tri}. We conclude in Sec.\;\ref{sec_Disc} with further discussions.

\section{Quantum Clocks in Quasi-Single Field Inflation}
\label{sec_QC}

In this section, we review the model of QSFI and its clock signals in the bispectrum, i.e., the 3-point function of the inflaton, taking into account general sound speed for the inflaton field as well as the massive spectator field. We use the spatially flat FRW metric $\di s^2=-\di t^2+a^2(t)\di\mb x^2=-a^2(\tau)(-\di\tau^2+\di\mb x^2)$ and work with both the comoving time $t$ and the conformal time $\tau$. The scale factor $a$ can be represented in both ways, $e^{H t}\simeq a \simeq -1/(H\tau)$, where $t$ ranges from $-\infty$ to $+\infty$ and $\tau$ takes values from $-\infty$ to 0. The time derivative of a quantity $\phi$ with respect to comoving $t$ will be denoted by a dot, $\dot\phi$, while the derivative with respect to conformal $\tau$ will be denoted by a prime, $\phi'$. These derivatives are related by, for example, $\phi'=a\dot \phi$.

In general, each point on the inflaton trajectory should be close to a local minimum in all transverse directions in the space of scalar fields. Consequently, if we expand the scalar potential around the inflaton trajectory, the terms quadratic in all transverse directions should take the form of a positive-definite mass matrix. Broadly speaking, an inflation scenario can be classified as QSFI if there is one or more mass eigenvalues $m_i\sim H$ in the transverse mass matrix, but there is no mass eigenvalue with $m_i\ll H$. It is possible to start with this classification, and write down directly an effective Lagrangian that describe the inflaton, the light massive fields with $m_i\sim H$, and their mutual interactions.

On the other hand, it is also possible to build QSFI models in a top-down approach, with the original model of the QSFI as a typical example \cite{Chen:2009we,Chen:2009zp}.
It has been shown that such a model may be naturally realized in the context  of the Standard Model \cite{Kumar:2017ecc} or supersymmetric theories \cite{Baumann:2011nk}.
In that case, the model consists of a pair of real scalar fields $(\si,\theta)$ which form a set of polar coordinates in the flat 2-dimensional field space. Then, the Lagrangian describes the model take the following form,
\bge
\label{SQSFI}
  S=\int\di^4x\sqrt{-g}\bigg[-\FR{1}{2}(\wt{R}+\si)^2(\pd_\mu\theta)^2-\FR{1}{2}(\pd_\mu\si)^2-V_{\text{sr}}(\theta)-V(\si)\bigg],
\ede
where we identify $\theta$ to be the inflaton field and thus the potential $V_\text{sr}(\theta)$ can take an arbitrary slow-roll form. On the other hand, we take $V(\si)$ such that $\si$ obtains a classical constant background $\si_0$ which is independent of time. After expanding the fields around their classical solutions $\theta_0=\theta_0(\tau)$ and $\si_0$, the Lagrangian for the fluctuation field has the following form,
\begin{align}
\label{ldcl}
\ld_\text{cl}=&\,\FR{a^2}{2}\Big[(\de\phi')^2-(\pd_i\de\phi)^2+(\de\si')^2-(\pd_i\de\si)^2\Big]-\FR{a^4m^2}{2}\de\si^2\n\\
&~+a^3\ka_1\de\si\de\phi'+a^3\ka_2\de\si^2\de\phi'+a^2\Big(\ka_3\de\si+\ka_4\de\si^2\Big)\Big[(\de\phi')^2-(\pd_i\de\phi)^2\Big]\n\\
&~-a^4\Big(\FR{\lam_3}{6}\de\si^3+\FR{\lam_4}{24}\de\si^4+\cdots\Big),
\end{align}
where we have defined $\de\phi=R\de\theta$ and $R\equiv \wt R +\si_0$, and the scale factor $a(\tau)\simeq -1/(H\tau)$. The first line of the above Lagrangian can be identified as free part $\ld_0$, with a massless scalar $\de\phi$ and a massive scalar $\de\si$ of mass $m^2=V''(\si_0)-\dot\theta_0^2$. In the second line, we have interactions with two-point derivative mixing between $\de\phi$ and $\de\si$, with coupling strength $\ka_1=2\dot\theta_0$. In addition, we have interaction terms derived from the kinetic term of $\theta$ in (\ref{SQSFI}), with the couplings $\ka_2=\dot\phi_0/R^2$, $\ka_3=1/R$, $\ka_4=1/(2R^2)$. In the third line, we have self-interactions of $\de\si$, with couplings $\lam_3=V'''(\si_0)$ and $\lam_4=V^{(4)}(\si_0)$. The self-interactions of $\de\theta$ derived from $V_\text{sr}(\theta)$ is suppressed by slow-roll parameters and will be neglected in the following.

The Lagrangian (\ref{ldcl}) can be generalized slightly by including general sound speeds for both $\de\phi$ and $\de\si$. It is straightforward to include non-unit sound speeds in the above top-down model. For instance, a new dim-6 operator $\frac{1}{2}\sqrt{-g}(\pd_\mu\theta\pd^\mu\si)^2/\Lambda^2$ in the original action (\ref{SQSFI}) would induce a term $\frac{1}{2}a^2(\dot\theta_0/\Lambda)^2\de\si'^2$ when evaluated on the inflation background, and after normalizing the time-derivative term of $\de\si$, we find a non-unit sound speed $c_\si^2=[1+(\dot\theta_0/\Lambda)^2]^{-1}$. But we will take the EFT approach at this point, and treat the sound speeds for both $\de\phi$ and $\de\si$ as arbitrary parameter. Therefore, we shall consider the following Lagrangian instead,
\begin{align}
\label{ldcl2}
\ld_\text{cl}=&\,\FR{a^2}{2}\Big[(\de\phi')^2-c_\phi^2(\pd_i\de\phi)^2+(\de\si')^2-c_\si^2(\pd_i\de\si)^2\Big]-\FR{a^4m^2}{2}\de\si^2\n\\
&~+a^3\ka_1\de\si\de\phi'+a^3\ka_2\de\si^2\de\phi'+a^2\Big(\ka_3\de\si+\ka_4\de\si^2\Big)\Big[(\de\phi')^2- (\pd_i\de\phi)^2\Big]\n\\
&~-a^4\Big(\FR{\lam_3}{6}\de\si^3+\FR{\lam_4}{24}\de\si^4+\cdots\Big) ~,
\end{align}
where we have normalized the field $\phi$ and $\si$ so that the coefficients of $\de\phi'$ and $\de\si'$ are canonical.
In principle there would also be corresponding corrections to the interactions in the second line of the above expression. For example, one can add a non-unit coefficient to the last term $(\pd_i\de\phi)^2$ in the second line, which can in principle different from $c_\phi^2$. But we won't consider such corrections further since most of $\ka_i$-interactions are irrelevant to our discussions expect for $\ka_1$, as we shall see below.

As noted in previous section, the QSFI model (\ref{ldcl2}) is interesting in several ways. Firstly, it can generate observably large non-Gaussianity due to the self-interaction of $\de\si$ and the bilinear mixing between $\de\si$ and $\de\phi$. The key point here is that the self-interactions of $\de\si$, i.e., the $(\lam_3,\lam_4)$ terms in (\ref{ldcl2}) are not constrained by slow-roll condition, and thus can be large in principle.
Secondly, the 3-point function $\la\de\phi(k_1)\de\phi(k_2)\de\phi(k_3)\ra$, contributed from $\lam_3$ term becomes an oscillatory function of $k_1/k_2$ in the $k_1\ll k_{2,3}$ limit, which goes like $B(k_1,k_2,k_3)\sin[\wt\nu\log(k_1/k_2)]$ when $m_\si>\frac{3}{2}H$, where $B(k_1,k_2,k_3)$ is non-oscillatory background and $\wt\nu=\sqrt{(m_\si/H)^2-9/4}$. Therefore, a measurement of $\wt\nu$ will reveal the mass $m_\si$ of the massive field $\de\si$, which is a key ingredient of the cosmological collider program. Thirdly, the appearance of the logarithm inside the sinusoidal function is actually due to the exponential expansion of the spacetime $a(t)\simeq e^{H t}$. In fact, it can be shown in general \cite{Chen:2015lza,Chen:2016qce} that the oscillation pattern takes the form of $\sin(a^{(-1)}(k_1/k_2))$ where $a^{(-1)}$ is the inverse function of the scale factor $a=a(t)$. Therefore, the squeezed bispectrum, namely the 3-point function of $\de\phi$ in the limit of $k_1\ll k_{2,3}$, can serve as a direct probe of the evolution history of the primordial universe.

\subsection{Diagrammatics and Mixed Propagator}

Next, we will review briefly the calculation of the bispectrum in the QSFI, but with the generalized Lagrangian (\ref{ldcl2}) describing the possibility of non-unit sound speeds. We follow the diagrammatic approach reviewed in \cite{Chen:2017ryl} and refer the readers to \cite{Chen:2017ryl} for a more comprehensive review of the formalism. Here we will only mention that the diagrammatic method presented in \cite{Chen:2017ryl} is essentially the same with the usual Feynman-diagram calculation, except that the space and the time are treated unequally, and that each vertex of the diagram comes with two types, labeled by either $+$ or $-$. Consequently, every internal vertex is associated with conserved 3-momentum flow and a time integration, and the propagator comes with four types, depending on the type of its two end points, and we should sum over diagrams with all possible type-assignment for all vertices.

In our space-time-asymmetric language, we Fourier transform the spatial directions but not the temporal direction. As a result, the propagator for a general scalar field $\varphi$ would depends on the time variables of both ends, together with the 3-momentum it carries. To derive the propagator, one first Fourier-decomposes the scalar field $\varphi$ in terms of its mode function $u(\tau,k)$,
\bge
\label{phimode}
  \varphi (\tau,\mb x)=\int\FR{\di^3\mb k}{(2\pi)^3}\Big[u (\tau, k)b (\mb k)+u^*(\tau, k)b^\dag(-\mb k)\Big]e^{\ii\mb k\cdot\mb x}.
\ede
Then the equation of motion $\varphi$ can be rewritten as an equation of $u(\tau,k)$. For instance, the mode function $u(\tau,k)$ of a scalar $\varphi$ with mass $m$ and sound speed $c_s$ satisfies the following equation,
 \begin{equation}\label{eom}
u''(\tau,k)-\frac{2}{\tau}u'(\tau,k)+\bigg(c_s^2 k^2+\frac{m^2}{H^2\tau^2}\bigg)u(\tau,k)=0.
\end{equation}
The solution that has the correct behavior (normalized coefficient and positive frequency) in the early time (flat spacetime) limit is uniquely determined to be,
\begin{align}
\label{modems}
  u (\tau,k)=-\FR{\ii\sqrt{\pi}}{2}e^{\ii\pi(\nu /2+1/4)}  H(-\tau)^{3/2}\text{H}_{\nu}^{(1)}(-c_s k\tau),
\end{align}
where $\nu\equiv\sqrt{9/4-(m/H)^2}$ which is real when $0\leq m\leq 3H/2$ and purely imaginary when $m>3H/2$, and $\text{H}_\nu^{(1)}(z)$ is the Hankel function of the first kind. In particular, when the mass $m=0$, the mode function becomes,
\bge
\label{modeml}
    u (\tau,k)=\FR{H}{\sqrt{2c_s^3k^3}}(1+\ii c_s k \tau)e^{-\ii c_s k\tau},
\ede

From the mode function, we can directly write down the four types of in-in propagators for the scalar field $\varphi$,
\bgs
\label{SKprop}
\begin{align}
G_{++}(k;\tau_1,\tau_2)=&~u(\tau_1,k)u^*(\tau_2,k)\theta(\tau_1-\tau_2)+u^{*}(\tau_1,k)u(\tau_2,k)\theta(\tau_2-\tau_1),\\
G_{+-}(k;\tau_1,\tau_2)=&~u^{*}(\tau_1,k)u(\tau_2,k),\\
G_{-+}(k;\tau_1,\tau_2)=&~u(\tau_1,k)u^*(\tau_2,k),\\
G_{--}(k;\tau_1,\tau_2)=&~u^{*}(\tau_1,k)u(\tau_2,k)\theta(\tau_1-\tau_2)+u(\tau_1,k)u^*(\tau_2,k)\theta(\tau_2-\tau_1),
\end{align}
\eds
where $\theta(z)$ is the Heaviside step function.

Given the propagator (\ref{SKprop}) and the mode functions (\ref{modems}) and (\ref{modeml}), one can follow the diagrammatic rules, which is essentially the same with the usual Feynman rules as reviewed in \cite{Chen:2017ryl}, and calculate the desired 3-point function of $\de\phi$ with intermediate exchange of $\de\si$. One highlight of QSFI is that the trilinear coupling $\lam_3\de\si^3$ in (\ref{ldcl2}) is not suppressed by the slow-roll conditions and can in principle be large so long as the theory remains perturbative. This is in contrast to the slow-roll-suppressed self-couplings of the inflaton $\de\phi$. Therefore, a 3-point function of the $\de\phi$ with cubic self-interaction of $\de\si$ would be the leading contribution to the bispectrum in the model, as we shall show more explicitly in the next subsection. For the moment, we only mention that the contribution from this cub self-coupling can be represented diagrammatically as,
\begin{eqnarray}
\label{3ptdiag}
\parbox{40mm}{\includegraphics{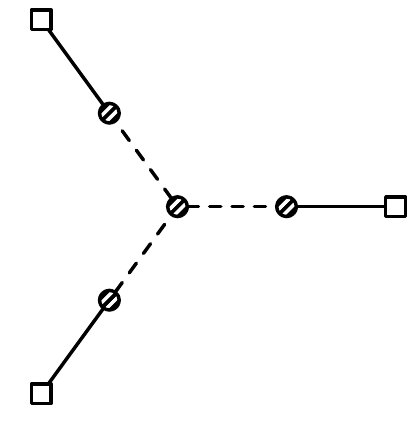}}~~~~=~~
\parbox{40mm}{\includegraphics{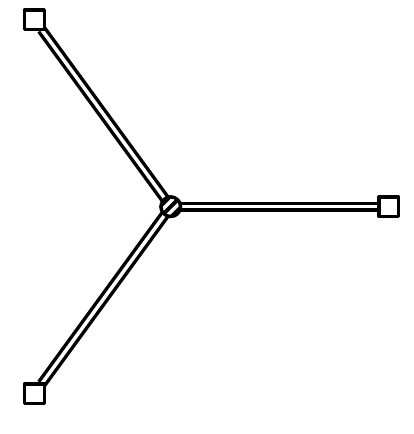}}
\end{eqnarray}
In the above equation, we have followed the diagrammatic notations introduced in \cite{Chen:2017ryl}. On the left-hand side, the solid lines represent the propagators of $\de\phi$ while dashed lines represent the propagator of $\de\si$. The squares at the three ends represent the endpoints of external lines, which are to be put to future infinity, while the shaded circles represent internal vertices. Each of the internal circles can be either $+$ type or $-$ type, with a corresponding sign associated with the coupling in the Feynman rule. Finally, we should sum over all possible $\pm$ types for all internal vertices. With 4 internal vertices, the left-hand side of (\ref{3ptdiag}) actually represents the sum of 16 diagrams.

As shown in \cite{Chen:2017ryl}, the calculation can be greatly simplified by noting that there is a repetitive structure on the left-hand side of (\ref{3ptdiag}), namely the bilinear mixing between $\de\phi$ and $\de\si$, and that one can first compute the bilinear mixing and then evaluate the whole diagram based on the bilinear mixing. The bilinear mixing is called a mixed propagator and is represented by the double-lines on the right-hand side of (\ref{3ptdiag}). This procedure not only reduces the number of diagrams to be computed, but also reduces the layers of integrals to be evaluated in each diagram. We refer the readers to \cite{Chen:2017ryl} for more details and now we will generalize the mixing propagator in \cite{Chen:2017ryl} to the case with general sound speed.

For clarity, we use $G_{ab}(k;\tau,\tau')$ to represent the propagator of $\de\phi$ and $D_{ab}(k;\tau,\tau')$ for $\de\si$, with $a,b=\pm$. Then, according to the diagrammatic rule, the mixed propagator is,
\begin{align}\label{mixedpropformal}
\mathcal G_\pm (k;\tau)&= \ii\ka_1\int_{-\infty}^0 \frac{\textmd{d}\tau'}{(-H\tau')^3}\left[D_{\pm+}(k;\tau,\tau')\partial_{\tau'}G_+(k ;\tau')-D_{\pm-}(k;\tau,\tau')\partial_{\tau'}G_-(k ;\tau')\right]\n\\
&=\frac{\pi \ka_1 H}{8 k^3 c_\phi c_\si^2} I_\pm (-c_\si k \tau ),
\end{align}
where the function $I_\pm(z)$ is given by
\begin{align}\label{resprop}
I_\pm(z)
=&~ e^{-\pi\textmd{\,Im\,}\nu} z^{3/2}\bigg\{2\,\text{Im}\bigg[ \text{H}_\nu^{(1)}(z)\int_0^\infty \frac{\textmd{d}z'}{\sqrt{z'}}\text{H}_{\nu^*}^{(2)}(z')e^{-\ii r z'}\bigg]\n\\
&~+\ii\text{H}_\nu^{(1)}(z)\int_0^z \frac{\textmd{d}z'}{\sqrt{z'}}H_{\nu^*}^{(2)}(z')e^{\mp \ii r z'}-\ii\text{H}_{\nu^*}^{(2)}(z)\int_0^z \frac{\textmd{d}z'}{\sqrt{z'}}H_{\nu}^{(1)}(z')e^{\mp \ii rz'}\bigg\}.
\end{align}
Here and in the following we use $r=c_\phi/c_\si$ to denote the ratio between the sound speed of $\de\phi$ and that of $\de\si$. For the integral in the first line of (\ref{resprop}) to converge, we give $c_\phi$ a small imaginary part $-\ii \epsilon$. The function $I_\pm(z)$ can be simplified a bit further into the following form,
\begin{equation}\label{respropdecomp}
I_\pm(z)= e^{-\pi\textmd{\,Im\,}\nu} z^{3/2}\Big\{ \text{H}_\nu^{(1)}(z)\big[e^{-\ii\pi(1-2\nu)/4}C_\nu^{r*} +B_{\nu\mp}^{r}(z)\big]+\text{H}_{\nu^*}^{(2)}(z)\big[e^{+\ii\pi(1-2\nu)/4}C_\nu^{r}+B_{\nu\pm}^r(z)\big]\Big\},
\end{equation}
where $C_\nu^r$ is a constant given by
\begin{align}\label{cvrescale}
C_\nu^r=\frac{2^{\nu }r^{\nu -1/2}}{\pi} &\;  \Gamma \Big(\frac{1}{2}-\nu \Big) \Gamma (\nu ) \, _2F_1\Big(\frac{1}{4} -\frac{ \nu}{2}
   ,\frac{3}{4}-\frac{\nu}{2};1-\nu ;\frac{1}{r^2}\Big) +(\nu \to -\nu) ,
\end{align}
while $B_{\nu\pm}^r(z)$ is a function given by the following integral for which we are not aware of a special-function representation
\begin{equation}
\label{frescale}
  B^r_{\nu\pm}(z)=-\ii\int_0^z\frac{\textmd{d}z'}{\sqrt{z'}}\text{H}_\nu^{(1)}(z') e^{\mp  \ii r z'}.
\end{equation}
We note that $C_\nu^{r=1}=\sqrt{2\pi}/\cos(\pi\nu)$.

In this following section, we shall examine various soft limit of the $n$-point functions which involve small momentum limit of the mixed propagator. Therefore, it is useful to write down an expression for $I_\pm(z)$ in the small $z$ limit,
\bge
\label{Ipmsmallz}
I_\pm (z)=2z^{3/2} C_\nu^r\bigg[\FR{\sin(\frac{\pi}{4}+\frac{\pi\nu}{2})}{{\Gamma (1+\nu)\sin (\pi\nu)}} \Big(\FR{z}{2}\Big)^\nu +(\nu\to-\nu)\bigg]+\order{z^2}.
\ede

\begin{figure}[tbph]
\centering
\includegraphics[width=0.55\textwidth]{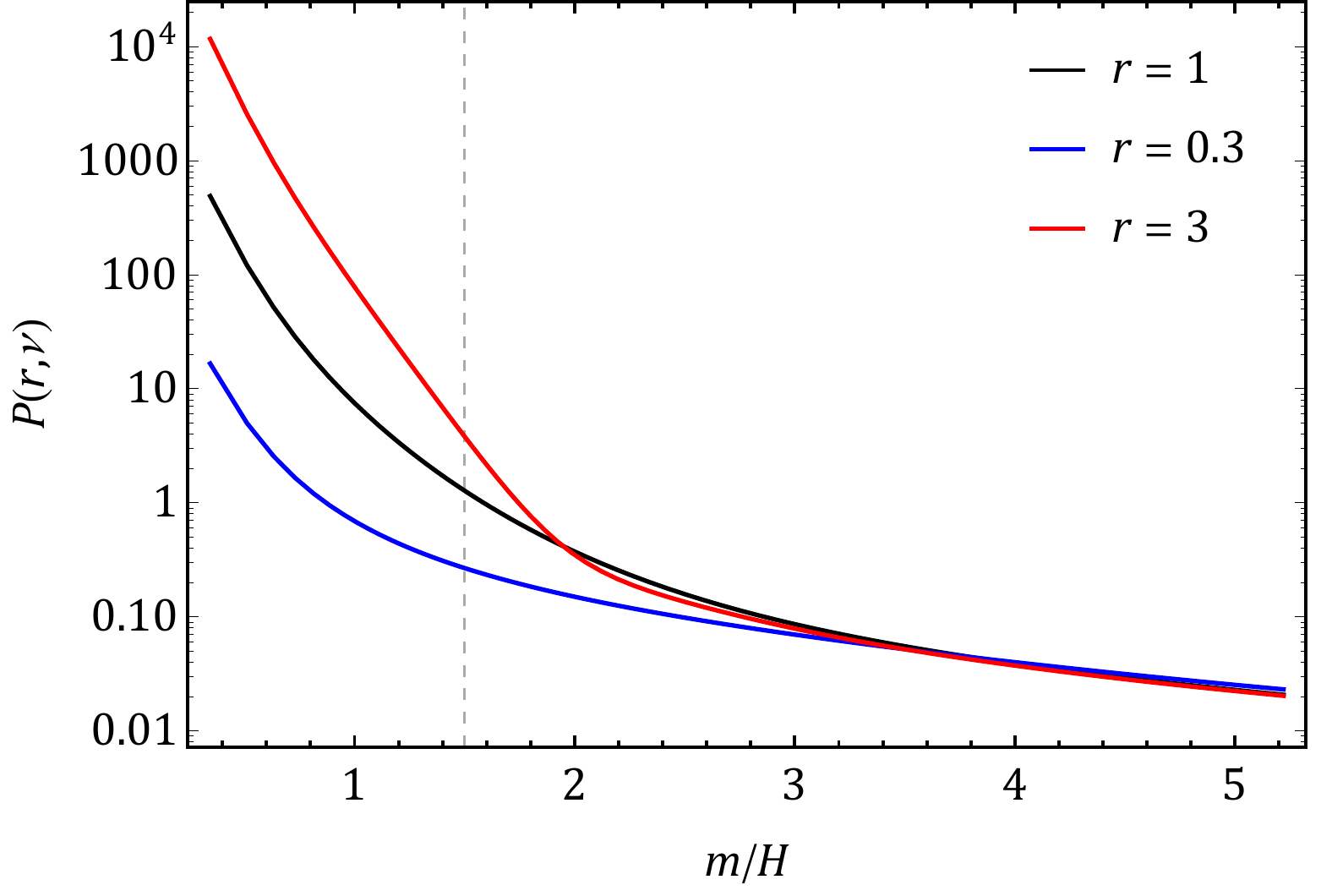}
\caption{The function $\mathcal{P}(r,\nu)$ in the leading-order correction to the power spectrum defined in (\ref{pnu}).  The blue, black, and red points correspond to sound-speed ratio $r=c_\phi / c_\sigma = 0.3,1,3$, respectively.}
\label{Fig_2pt}
\end{figure}

As an immediate application of the mixed propagator (\ref{mixedpropformal}), we calculate the two-point function of the inflaton field $\de\phi$. Using the diagramatic rule, we have, to the first order in $\ka_1$,
\begin{align}
\label{2ptrescale}
\la \delta \phi \delta \phi\ra'=&\parbox{85mm}{\includegraphics[width=85mm]{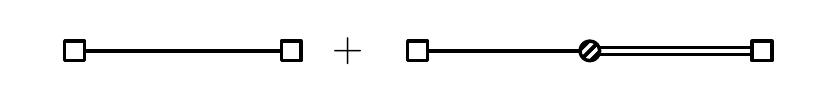}}\n\\
=&~G_+(k;\tau=0)+i\ka_1 \int_{-\infty}^0 \frac{\textmd{d}\tau}{(-H\tau)^3}\Big[ \mathcal G_+(k;\tau)\partial_{\tau}G_+(k;\tau) - \mathcal G_-(k;\tau)\partial_{\tau} G_-(k;\tau)\Big]\n\\
 =&~\frac{H^2}{2k^3 c_\phi^3 }\bigg[1+ \Big(\FR{\ka_1 }{H}\Big)^2  \mathcal P (r,\nu)\bigg],
\end{align}
where $\mathcal P (r,\nu)$ is defined as
\begin{equation}\label{pnu}
\mathcal P (r,\nu)\equiv \frac{\pi r}{4}\,\textmd{Im}\,\bigg[\int_0^\infty\frac{\textmd{d}z}{z^2}\,e^{-irz}I_+(z) \bigg].
\end{equation}
We plot $\mathcal{P}(r,\nu)$ for different $r$ in Fig.\;\ref{Fig_2pt}.

The correction to the power spectrum from the intermediate $\de\si$ field is an overall scaling of the leading order result, i.e., the first diagram on the right-hand side of (\ref{2ptrescale}). This overall shift of the power spectrum is not directly observable, and thus is not interesting for our purpose.\footnote{However, the correction is important if one would like to use the $n_s$-$r$ diagram to pin down simple inflation models. Due to the rolling background of inflation, even Planck suppressed dimension-5 operators of the QSFI type may have an observable impact on the $n_s$-$r$ diagram\cite{Jiang:2017nou, Tong:2017iat}.} Due to the constraint from the scale invariance, the ``quantum clock signal'' in QSFI can show up only as a function of dimensionless combination of external momenta, which is absent at 2-point level. The first nontrivial example of a quantum clock, therefore, comes from the 3-point function of the inflaton, which we shall review in the next subsection.

\subsection{Bispectrum in QSFI with General Sound Speed}

With the help of the mixed propagator, we can now explore the $n$-point correlation functions in QSFI more readily. Before entering the trispectrum which is at the 4-point level, we shall first revisit the 3-point functions of QSFI in the rest of the subsection, taking account of general sound speeds for both the inflaton $\de\phi$ and the spectator field $\de\si$. The quantity we can calculate directly is the 3-point function $\la\de\phi(\mb k_1)\de\phi(\mb k_2)\de\phi(\mb k_3)\ra'$ of the inflaton field $\de\phi$, with the prime indicating that the momentum-conservation $\de$-function is stripped off. On the other hand, it is conventional to parameterize the 3-point function in terms of the dimensionless and scale-invariant shape function $S(k_1,k_2,k_3)$ as follows,
\begin{equation}\label{3ptshaperaw}
\begin{aligned}
\langle \zeta(\mathbf{k}_1) \zeta(\mathbf{k}_2) \zeta(\mathbf{k}_3) \rangle' = \left(2 \pi \right)^4P_\zeta^2  \frac{1}{(k_1k_2k_3)^2}S(k_1,k_2,k_3),
\end{aligned}
\end{equation}
where $\zeta =-H\de\phi/\dot\phi_0$ and $P_\zeta = H^4/(4\pi^2c_\phi^3\dot\phi_0^2)$ is the observed scalar power spectrum. Note that the appearance of the sound speed in this power spectrum expression may look unfamiliar (comparing with e.g.~that in \cite{Chen:2006nt}) but this is because for generality we have rescaled both $\phi_0$ and $\delta\phi$ by the same factor so that the $(\delta\phi')^2$ term in Eq.~\eqref{ldcl2} is canonical.

At the tree level, possible contributions to the 3-point function of $\de\phi$ include the following diagrams.
\begin{eqnarray}
\label{3ptdiag2}
\parbox{35mm}{\includegraphics[width=35mm]{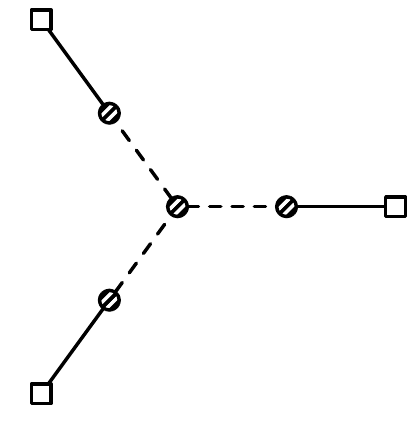}}\hspace{4mm}
\parbox{35mm}{\includegraphics[width=35mm]{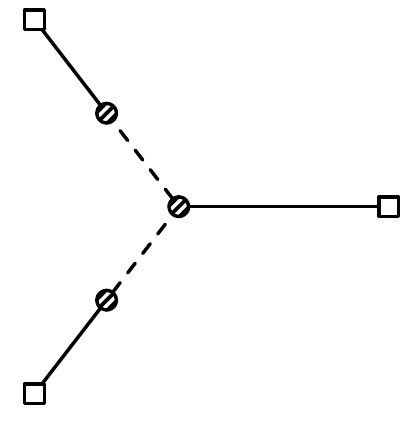}}\hspace{4mm}
\parbox{35mm}{\includegraphics[width=35mm]{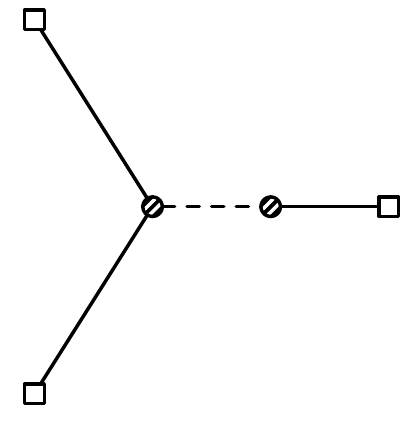}}\hspace{4mm}
\parbox{35mm}{\includegraphics[width=35mm]{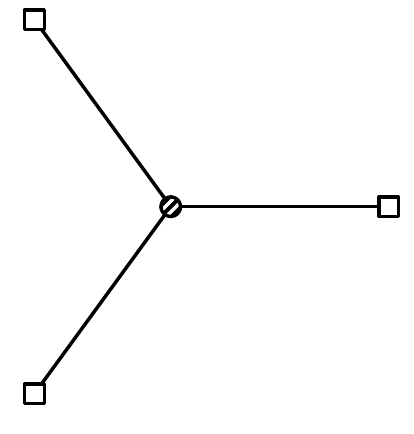}}
\end{eqnarray}
By dimension analysis of the couplings, it is not difficult to see that the first diagram parametrically dominates the 3-point function, if the contributions from the internal propagator can be assumed to be of $\order{1}$ (\footnote{This is no longer true when the internal propagator receives additional large suppression from a large mass $m$ or large enhancement from a small sound speed $c_\si$. But we will not consider these extreme cases in this simple estimate.}). In fact, the first diagram is proportional to $(\ka_1/H)^3(\lam_3/H)\sim (\ka_1/H)^3(m/H)$, the second one is $(\ka_1/H)^2\ka_2\sim\ka_1^3/H^2R$, the third one is $(\ka_1/H)(H\ka_3)\sim \ka_1/R$, and the last one has to be suppressed at least by $1/R^2$ since $\de\phi$ has to be derivative coupled in the slow-roll limit. Here $R$ can be regarded as the cut-off scale of the effective field theory of QSFI, at which all dimensionless effective couplings are of $\order{1}$ in a natural theory.

It is helpful at this point to see how large the cutoff scale $R$ is. In this paper we have been focusing on the weak coupling regime where the mixed propagator, i.e., the two-point correlation $\la\de\phi\de\si\ra$, is dominated by the diagram with one $\ka_1$-insertion, and diagrams with more $\ka_1$-insertions can be neglected. Note that each additional $\ka_1$-insertion introduces one more bilinear vertex proportional to $\ka_1$, as well as one more propagator of $\de\si$ which scales as $1/m$ when $m>3H/2$. Therefore, the weak coupling regime roughly corresponds to $\ka_1<m$. In addition, we are interested in the region where the quantum clock signal is observably large. In the following we shall see that the clock signal in the weak coupling regime is suppressed by a Boltzmann factor $\sim e^{-\pi m/H}$ when $m\gg H$ (\footnote{In the strong coupling regime, the Boltzmann factor is given instead by $e^{-\pi\sqrt{m^2+\ka_1^2}/H}$ \cite{An:2017hlx, An:2017rwo}. But this does not change our estimate here and the visibility of the clock signal still requires that $\max\{\ka_1,m\}\lesssim H$. }). Therefore, it is preferable that $m$ is greater than $3H/2$ but not too heavier than $H$. Consequently, we are mostly interested in the region when $\ka_1\leq H$. Recall that $\ka_1=2\dot\phi_0/R$ in the model described in (\ref{ldcl}), we see that the weak coupling plus the large clock signal requires $R>\dot\phi_0/H\sim P_\zeta^{-1/2}H$. Given that $P_\zeta^{-1/2}\sim 10^5$, we see that $R$ is indeed a very large scale and a diagram with an additional $1/R$ factor is indeed very much suppressed. This further justifies the above estimate of contributions from different diagrams.

Therefore we shall only consider the first diagram of (\ref{3ptdiag2}) in the following. Using the diagrammatic rule in \cite{Chen:2017ryl}, the first diagram can be calculated as follows,
\begin{align}
\label{3ptraw}
\la\delta \phi (\mathbf{k}_1 ) \delta \phi (\mathbf{k}_2 ) \delta \phi (\mathbf{k}_3 )\ra'=&~2\lambda_3
\text{Im} \int_{-\infty}^0 \frac{d\tau}{(-H\tau)^4}\mathcal G_+(k_1;\tau)\mathcal G_+(k_2;\tau)\mathcal G_+(k_3;\tau) \n\\
 =&~\frac{\pi^3\ka_1^3 \lambda_3 }{256 H k_1^3 k_2^3 c_\phi^3 c_\sigma^3}
\text{Im} \int_{0}^\infty \frac{dz}{z^4}I_+ \Big(\frac{k_1}{k_3}z\Big)I_+\Big( \frac{k_2}{k_3}z\Big)I_+ (z).
\end{align}

We are more interested in the squeezed limit $k_1\ll k_2\simeq k_3$, because this is the configuration that the quantum clock signals show up. For heavy enough $\de\si$, $m>3H/2$, the shape function $S(k_1,k_2,k_3)$ contain the clock signal in the squeezed limit as an oscillating function of $k_1/k_3$. To see the oscillation more explicitly, we expand the above 3-point function (\ref{3ptraw}) in the small $k_1/k_3$ limit, using the small-$z$ expansion of the mixed propagator (\ref{Ipmsmallz}). The result can be expressed in terms of the shape function $S(k_1,k_2,k_3)$ introduced in (\ref{3ptshaperaw}), as,
\begin{align}\label{3ptshape}
 S(k_1,k_2,k_3)
 \simeq  \FR{1}{\sqrt{P_\zeta}} \bigg( \frac{\ka_1}{H} \bigg)^3 \bigg( \frac{\lambda_3}{H} \bigg) \FR{1}{c_\phi^{3/2}}\,\text{Im}\, \bigg[ s_+ \bigg(  \frac{k_1}{k_3} \bigg)^{1/2+\nu} + s_- \left( \frac{k_1}{k_3} \right)^{1/2-\nu} + \mathcal{O} \bigg(  \frac{k_1}{k_3} \bigg)\bigg],
\end{align}
where $\nu=\sqrt{9/4-(m/H)^2}$ and the coefficients $s_{\pm,0}$ are,
\begin{align}\label{3ptreform}
s_\pm &= -\frac{\pi^2 r^3 C_\nu^r \sin \big(\frac{\pi }{4}\pm\frac{\pi  \nu }{2}\big) }{2^{8\pm\nu } \Gamma (1\pm\nu)\sin(\pm\pi\nu)}\int_{0}^\infty dz\,  I_+^2 (z) z^{-5/2\pm\nu} ,
\end{align}
where $I(z)$ is given in (\ref{respropdecomp}) and $C_\nu^r$ is given in (\ref{cvrescale}).

\begin{figure}[t]
\centering
\includegraphics[width=0.8\textwidth]{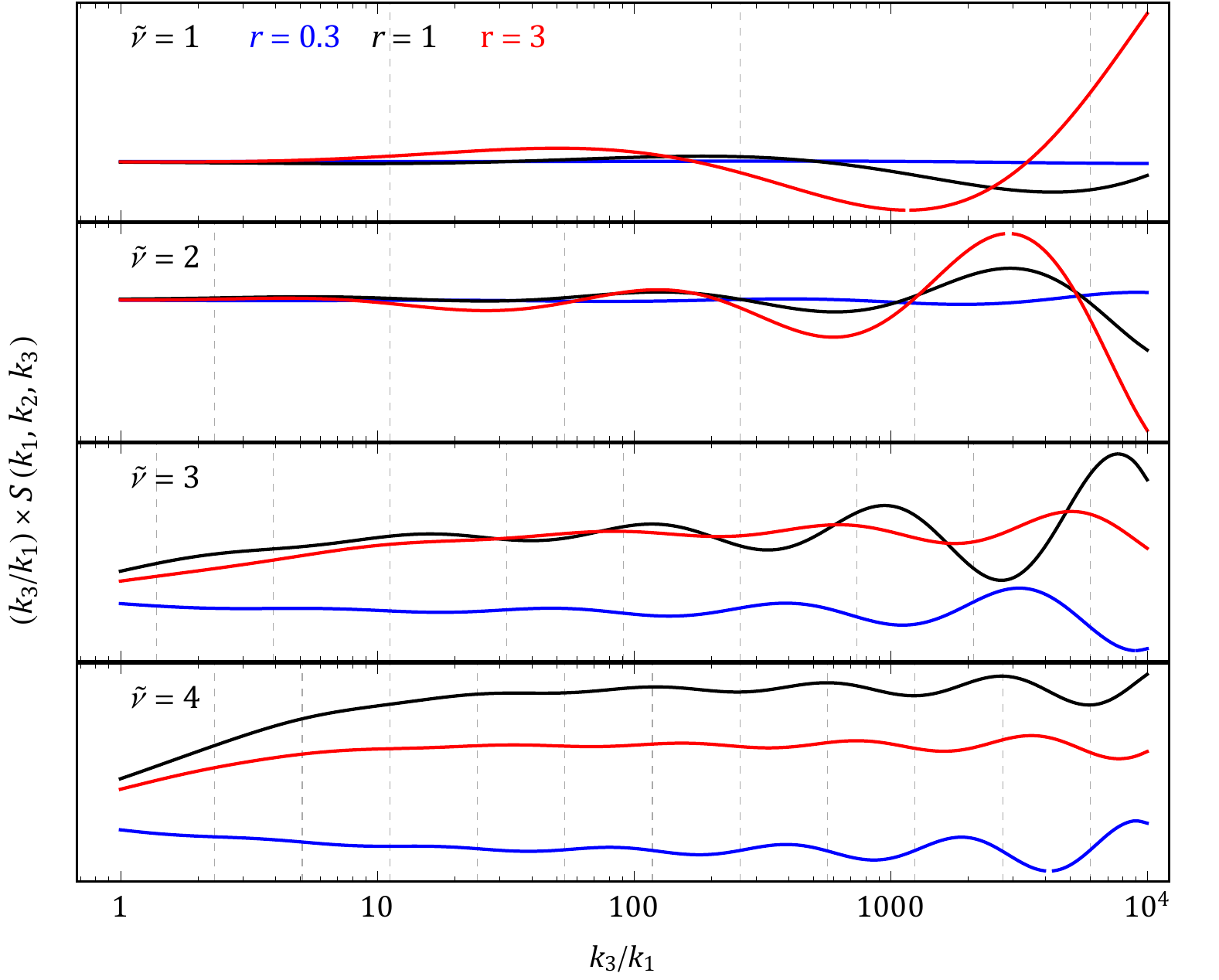}
\caption{The clock signals in the squeezed limit of the bispectrum in QSFI with general sound speed, as function of momentum ratio $k_3/k_1$, with $k_1$ the soft momentum. The four rows correspond to four choices of $\wt\nu=\sqrt{(m/H)^2-9/4}$ and, in each row, the blue, black, and red curves show the signals for sound-speed ratio $r=c_\phi / c_\sigma = 0.3,1,3$, respectively. }
\label{Fig_3ptSqueezed}
\end{figure}

\begin{figure}[t]
\centering
\includegraphics[width=0.49\textwidth]{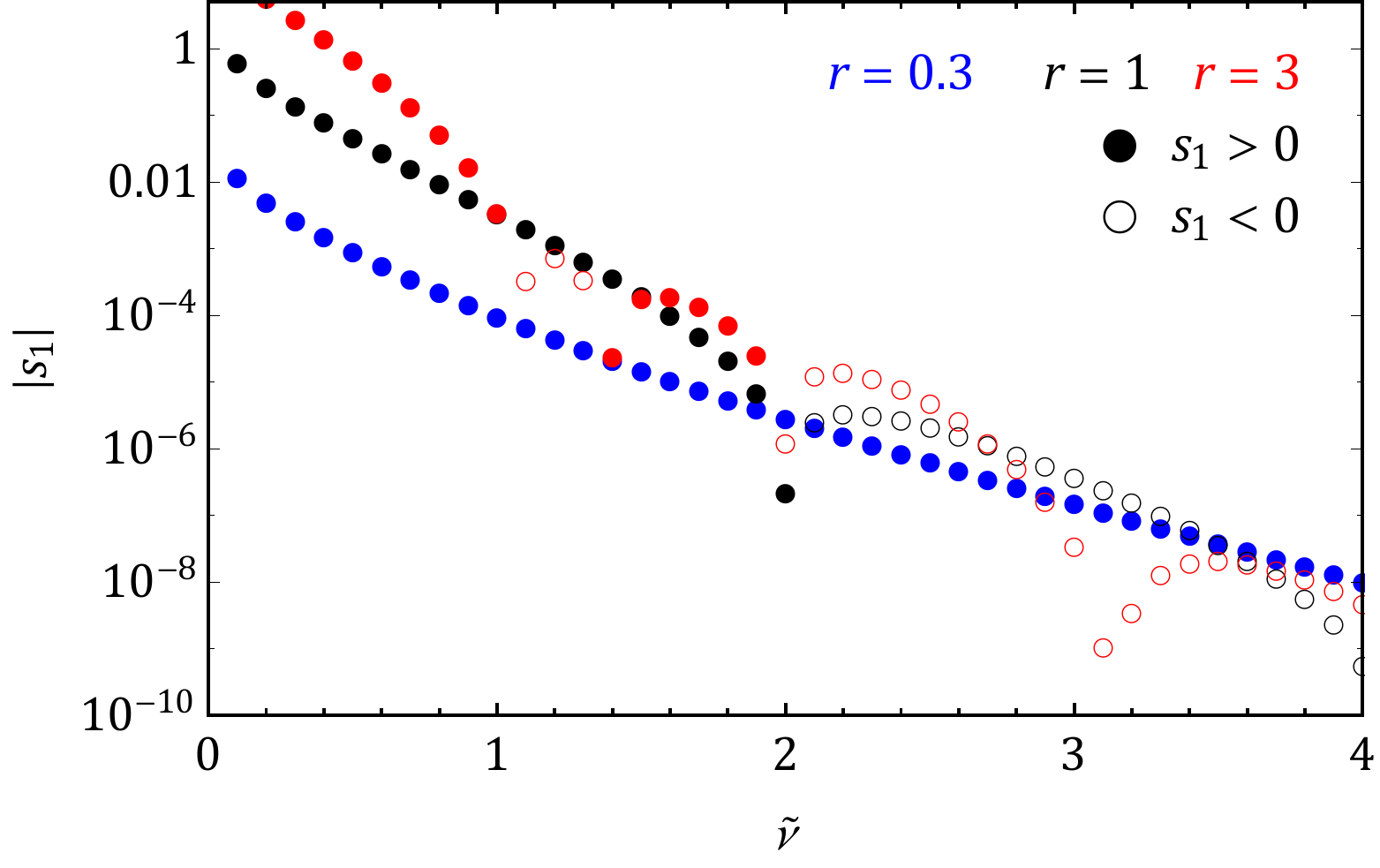}
\includegraphics[width=0.49\textwidth]{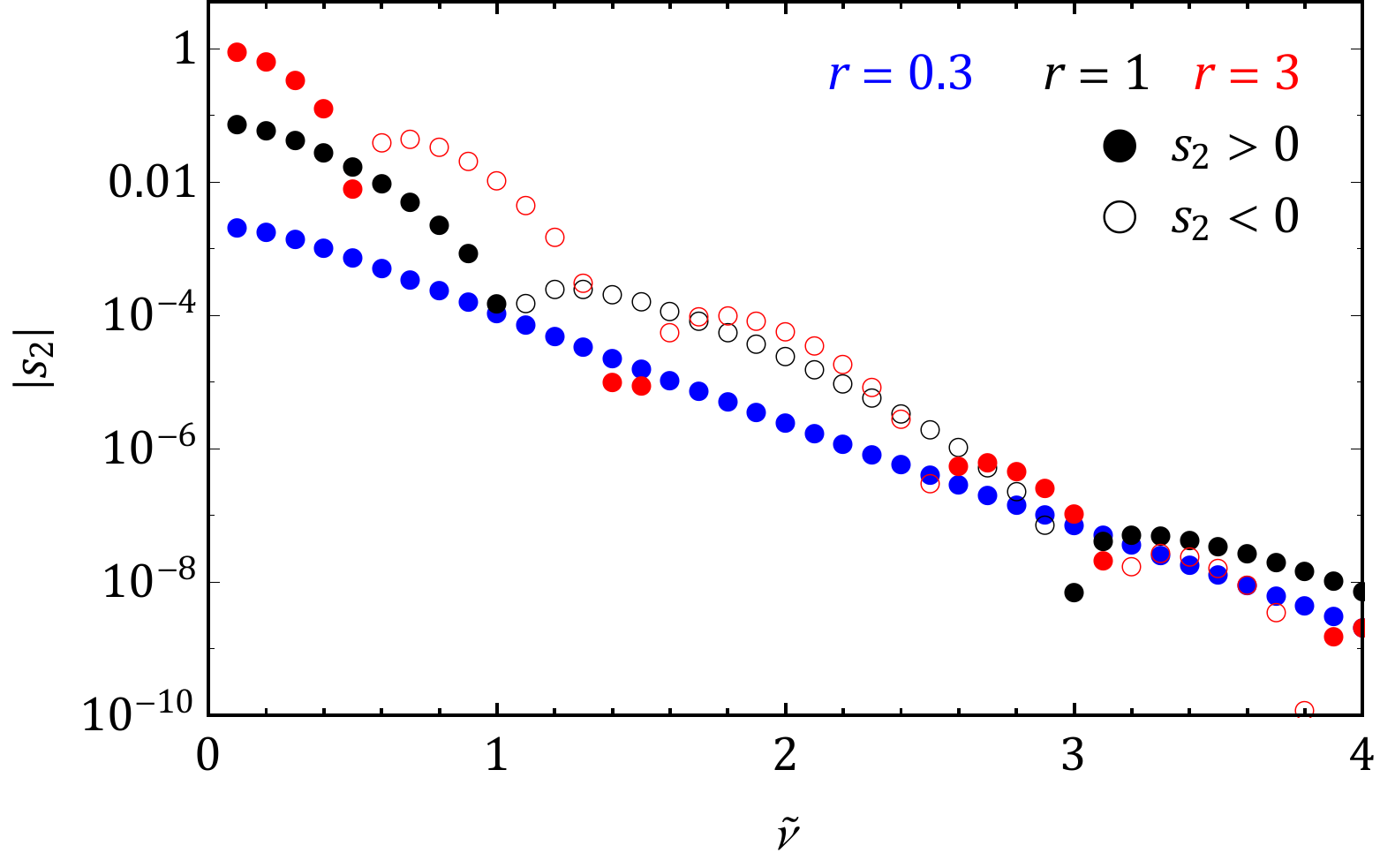}
\caption{The coefficient $s_1$ (left panel) and $s_2$ (right panel) for the oscillatory clock signal in the squeezed limit of the bispectrum. The solid dots (empty circles) denote positive (negative) $s_{1,2}$. The blue, black, and red points correspond to sound-speed ratio $r=c_\phi / c_\sigma = 0.3,1,3$, respectively. }
\label{Fig_scoef}
\end{figure}

It can be seen from (\ref{3ptshape}) that the shape function contains an oscillating piece in $s_\pm$-terms if $m>3H/2$ so that $\nu$ is pure imaginary. In this case,
\begin{align}
\label{Sclock}
 S(k_1,k_2,k_3)\simeq
  \frac{\pi^5}{64\sqrt{P_\zeta}}  \bigg( \frac{\ka_1}{H} \bigg)^3 \bigg( \frac{\lambda_3}{H} \bigg) \FR{1}{c_\phi^{3/2}}\Big(\FR{k_1}{k_3}\Big)^{1/2}\bigg[s_1\sin\Big(\wt\nu\log\FR{k_1}{k_3}\Big)+s_2\cos\Big(\wt\nu\log\FR{k_1}{k_3}\Big)+\cdots\bigg],
\end{align}
where $\wt\nu\equiv \text{Im}\,\nu$ which is positive when $m>3H/2$, and we have defined $s_{1}\equiv\Re(s_+ + s_-)$, $s_{2}\equiv\Im(s_- - s_+)$ and neglected non-oscillating and sub-leading terms. It is clear from (\ref{Sclock}) that the squeezed limit of the bispectrum contains an oscillating component as a function of momentum ratio $k_1/k_3$. The frequency $\wt\nu=\sqrt{(m/H)^2-9/4}$ is determined by the mass of the heavy scalar $\de\si$ and the logarithmic dependence is directly related to the exponential expansion of the universe during inflation. The strength of the oscillating piece is determined by the couplings $\lam_3$, $\ka_1$, the sound speeds $(c_\phi,c_\si)$, and also by the mass $m$ of $\de\si$ through the coefficients $s_{1,2}$. The result with unit sound speed $c_\phi=c_\si=1$ has been worked out in \cite{Chen:2017ryl}, and here it is generalized to the cases with the possibility of non-unit sound speeds. The new features of general sound speeds include not only an overall factor, namely $r^3c_\phi^{-3/2}$ in (\ref{Sclock}), but also, more interestingly, the corrections to coefficients $s_{1,2}$. As a result, both the amplitude and the phase of the clock signal, but not its frequency, are affected. We show in Fig.\;\ref{Fig_3ptSqueezed} the clock signals in the squeezed limit of the shape function $S(k_1,k_2,k_3)$ as functions of the momentum ratio $k_1/k_3$, for different choices of masses and sound speeds. It can be seen from the figure that the frequency of the oscillatory clock signal depends only on the mass $m$, via the dependence on $\wt\nu$ in (\ref{Sclock}), while different choices of sound-speed ratio $r$ do change the phase and the amplitude of the signals. To make this point clearer, we further show in Fig.\;\ref{Fig_scoef} the coefficients $s_{1,2}$ as functions of $\wt\nu$, for different choices of sound-speed ratio $r$. It can be seen from this figure that the coefficients $s_{1,2}$ drop exponentially when $\wt\nu$ gets large, as a result of the Boltzmann suppression.

\section{Quantum Clocks in the Trispectrum}
\label{sec_Tri}

In this section we calculate explicitly the leading 4-point function of the inflaton field $\de\phi$ in QSFI, and extract the quantum clock signals in two different soft limits of the trispectrum. We can again rewrite the 4-point function in terms of $\zeta$-modes via the relation $\zeta=-H\de\phi/\dot\phi_0$, and we can parameterize the 4-point function of $\zeta$-modes in terms of the dimensionless and scale-invariant shape function $T(\mb k_1,\mb k_2,\mb k_3,\mb k_4)
$, in the following way.
\begin{equation}\label{4ptdef}
\la\zeta(\mb k_1)\zeta(\mb k_2)\zeta(\mb k_3)\zeta(\mb k_4)\ra'=(2\pi)^6 P^3_\zeta \frac{K^3}{(k_1k_2k_3k_4)^3}T(\mb k_1,\mb k_2,\mb k_3,\mb k_4),
\end{equation}
where $K=k_1+k_2+k_3+k_4$. In the following, we shall write down an expression for the shape function. Then, in the next two subsections, we will extract quantum clock signals from this shape function in two types of soft limits.

Similar to the case of 3-point function, the leading contributions come from the self-interactions of $\de\si$. But in the case of 4-point function, we have two types of diagrams, one proportional to $\ka_1^4\lam_4$, the other proportional to $\ka_1^4\lam_3^2$, as shown below.
\begin{eqnarray}
\label{4ptdiag}
\parbox{105mm}{\includegraphics[width=105mm]{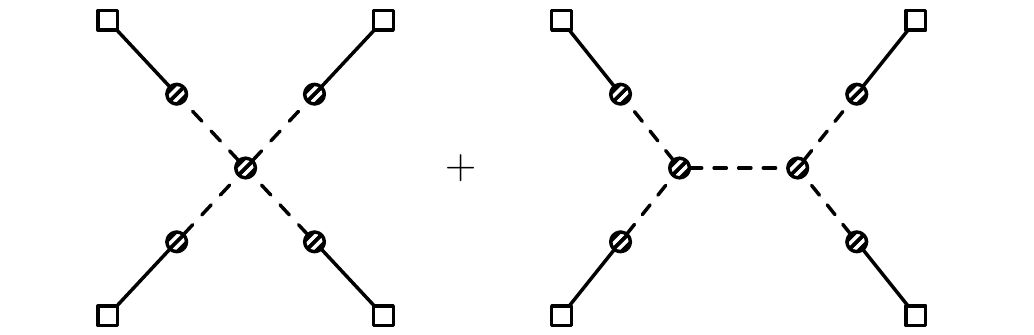}}\hspace{-5mm}+\text{2 permutations}.
\end{eqnarray}
In above diagrams only the $s$-channel of $\lam_3^2$-diagram is shown explicitly, and it is understood that one should also include the $t$ and $u$ channels.

The contact diagram with the quartic coupling is very similar to the $\lam_3$-diagram in the 3-point function, and it is easy to write down the corresponding result following the diagrammatic rule,
\begin{align}\label{4ptcontact}
&~\la \delta\phi( \textbf{k}_1)  \delta\phi( \textbf{k}_2) \delta\phi( \textbf{k}_3) \delta\phi( \textbf{k}_4)\ra_{c}'\n\\
=&~2\lambda_3
\text{Im} \int_{-\infty}^0 \frac{d\tau}{(-H\tau)^4}\mathcal G_+(k_1;\tau)\mathcal G_+(k_2;\tau)\mathcal G_+(k_3;\tau) \mathcal G_+(k_4;\tau)\n\\
=&~\frac{\pi^4 \ka_1^4 \lambda_4r^5}{2^{11}c_\phi^9}\FR{1}{(k_1k_2k_3k_4)^3} \, \textmd{Im}\int_0^{\infty}\frac{\textmd{d}z}{z^4}I_+(k_1z)I_+(k_2z)I_+(k_3z)I_+(k_4z).
\end{align}
On the other hand, the $s$-channel diagram with two cubic-vertices can be written down as follows,
\begin{align}\label{raw4pt}
&\la \delta\phi( \textbf{k}_1)  \delta\phi( \textbf{k}_2) \delta\phi( \textbf{k}_3) \delta\phi( \textbf{k}_4)\ra_{s}'\n\\
=&-\lam_2^4\sum_{a,b=\pm}\int_{-\infty}^0\FR{\di\tau_1\di\tau_2}{(H^2\tau_1\tau_2)^4} \mathcal{G}_a(k_1;\tau_1)\mathcal{G}_a(k_2;\tau_1)\mathcal{G}_b(k_3;\tau_2)\mathcal{G}_b(k_4;\tau_2)D_{ab}(k_s,\tau_1,\tau_2)\n\\
=&~\frac{\pi^5 \ka_1^4\lambda_3^2}{8192H^2}\frac{k_s^3}{(k_1k_2k_3k_4)^3}\frac{r^5}{c_\phi^9 }e^{-\pi\,\textmd{Im}\,\nu} \,\textmd{Re}\,\bigg[\int_{\text{T}\{zz'\}} \Big(J_{12s}^{+-}(z)J_{34s}^{++}(z')+  J_{34s}^{+-}(z)  J_{12s}^{++}(z')\Big) \n\\
&- \int_{zz'}J_{34s}^{+-}(z) J_{12s}^{-+}(z') \bigg].
\end{align}
where $k_s\equiv |\mb k_1+\mb k_2|$, and we have introduced the shorthand notation for the integral,
\begin{align}
&\int_{\text{T}\{zz'\}}\equiv\int_0^\infty\textmd{d}z \int_0^z \textmd{d}z',
&&\int_{zz'}\equiv\int_0^\infty \textmd{d}z\int_0^\infty \textmd{d}z'.
\end{align}
Furthermore, we have defined,
\begin{align}
  &J_{ij\ell}^{\pm +}(z)=z^{-5/2} I_\pm\Big(\frac{k_i}{k_\ell}z\Big)I_\pm\Big(\FR{k_j}{k_\ell}z\Big)\text{H}_{\nu}^{(1)}(z),
  &&J_{ij\ell}^{\pm -}(z)=z^{-5/2} I_\pm\Big(\frac{k_j}{k_\ell}z\Big)I_\pm\Big(\FR{k_j}{k_\ell}z\Big)\text{H}_{\nu^*}^{(2)}(z).
\end{align}
The expressions for $t$ and $u$-channels can be found similarly, by the substitution $(2\leftrightarrow 3, s\to t)$ and $(2\leftrightarrow 4, s\to u)$, respectively, where $k_t\equiv |\mb k_1+\mb k_3|$ and $k_u\equiv |\mb k_1+\mb k_4|$.

From the above results, we can find the shape function $T$, defined in (\ref{4ptdef}), as,
\begin{align}\label{4ptform}
 T=&~\FR{\pi^3c_\phi^3}{2^{15}P_\zeta} \Big(\frac{\ka_1}{H}\Big)^4\bigg[\frac{4 \lambda_4}{\pi}t_c+ \Big(\frac{\lambda_3}{H}\Big)^2\big(t_s+t_t+t_u\big)\bigg],
\end{align}
where $t_c$ and $t_s$ are,
\begin{align}\label{4pttcoef}
t_{c}=&~ r^5\,\textmd{Im}\,\int_0^{\infty}\frac{\textmd{d}z}{z^4}I_+\Big(\frac{k_1}{K} z\Big)I_+\Big(\frac{k_2}{K} z\Big)I_+\Big(\frac{k_3}{K} z\Big)I_+\Big(\frac{k_4}{K} z\Big),\\
t_s=&~e^{-\pi \,\textmd{Im}\, \nu}r^5\Big(\FR{k_s}{K}\Big)^3\,\textmd{Re}\,\bigg[\int_{\text{T}\{zz'\}}\Big(J_{12s}^{+-}(z)J_{34s}^{++}(z')+  J_{34s}^{+-}(z)  J_{12s}^{++}(z')\Big)\n\\
&~ -\int_{zz'} J_{34s}^{+-}(z) J_{12s}^{-+}(z') \bigg],
\end{align}
and $t_t$ and $t_u$ can be written down similarly by permutations.

Given (\ref{4ptform}), one can readily calculate the shape function $T$ for any momentum configuration. The mixed propagator renders the numerical evaluation quite easy. The only subtlety comes from the possible infrared divergences when $\nu$ is real. Although such divergences in any subgraph must cancel out in the final answer, the cancelation is difficult to implemented in numerical calculation. A simple way to circumvent this problem is to choose an integration contour with half-way Wick rotation, as described in \cite{Chen:2017ryl}.

We are mostly interested in the parameter space of this 3-parameter set where the quantum clock signal shows up. This happens whenever a propagator of $\de\si$ field in (\ref{4ptdiag}) goes soft, i.e., when its 3-momentum goes to zero. In this case, the corresponding massive field has much longer wavelength than the shorter modes, and its harmonic oscillations in the classical regime serves as a background clock for the shorter massless curvature modes, generating the quantum clock signals \cite{Chen:2015lza,Chen:2016qce}.

\subsection{Triangular Limit}

\begin{figure}[tbph]
\centering
\includegraphics[width=0.3\textwidth]{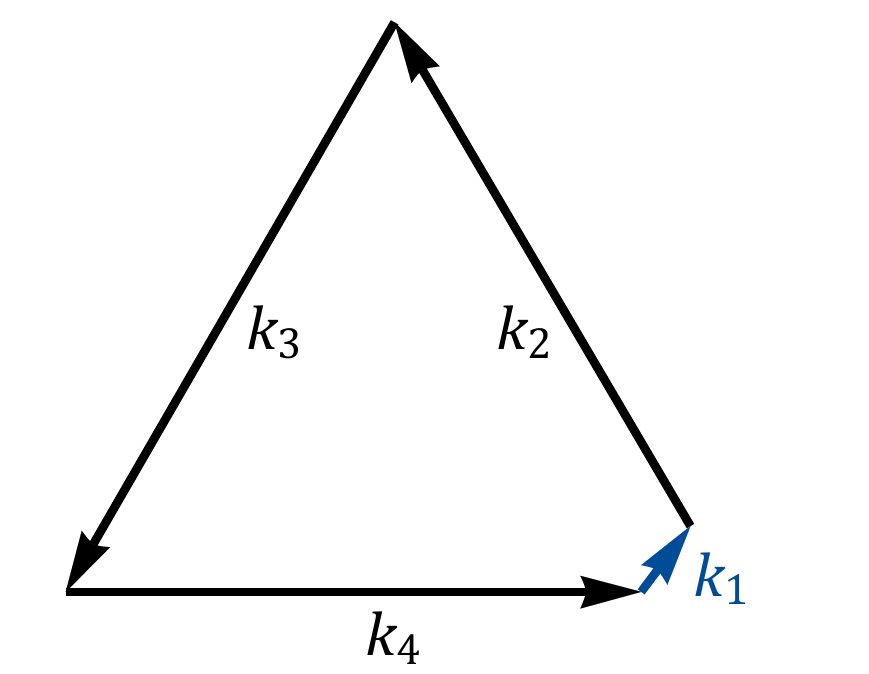}
\caption{The squeezed limit of the shape function.}
\label{Fig_Triangular}
\end{figure}

When one of the four external momenta go soft, the conservation of momentum requires that the other three momenta form a nearly closed triangle. Therefore we call such a soft limit the triangular limit. Without loss of generality, we can take $k_1\to 0$, as shown in Fig.\;\ref{Fig_Triangular}. In both of diagrams in (\ref{4ptdiag}), the soft momentum $k_1$ is carried by a mixed propagator $\mathcal{G}(k_1;\tau)$, and thus we expand it in the small $k_1$ limit, using (\ref{Ipmsmallz}). The shape function then becomes,
\begin{align}
  T=\FR{\pi^3c_\phi^3}{2^{15}P_\zeta} \Big(\frac{\ka_1}{H}\Big)^4\,\text{Re}\,\bigg[t_+\Big(\FR{k_1}{K}\Big)^{3/2+ \nu}+t_-\Big(\FR{k_1}{K}\Big)^{3/2-\nu}+\cdots\bigg],
\end{align}
where we retain only terms non-analytical in the soft momentum $k_1$, i.e., the two terms proportional to $(k_1/K)^{\pm\nu}$, and neglect analytical terms. The coefficients $t_\pm$ of the non-analytical terms are given by,
\begin{align}
  t_\pm =&~\FR{2^{1\mp\nu} r^5\sin(\frac{\pi}{4}\pm\frac{\pi\nu}{2})C_\nu^r}{ \Gamma (1\pm\nu)\sin (\pm\pi\nu)}\bigg[-\FR{4\ii\lam_4}{\pi}\mathcal{I}_c^\pm+\Big(\frac{\lambda_3}{H}\Big)^2\big(\mathcal{I}_s^\pm+\mathcal{I}_t^\pm+\mathcal{I}_u^\pm\big)\bigg] ,
\end{align}
and the four integrals $\mathcal{I}_{c,s,t,u}$ come from the four diagrams in (\ref{4ptdiag}), namely the contact diagram with quartic $\de\si$-coupling, the $s,t,u$-channel diagrams with two cubic $\de\si$-couplings, respectively. The integrals $\mathcal{I}_c$ and $\mathcal{I}_s$ are,
\begin{align}
\mathcal{I}_c^\pm=&~ \int_0^\infty\di z\,I_+\Big(\frac{k_2}{K} z\Big)I_+\Big(\frac{k_3}{K} z\Big)I_+\Big(\frac{k_4}{K} z\Big)z^{-5/2\pm \nu}\\
\mathcal{I}_s^\pm=&~\Big(\FR{k_2}{K}\Big)^{3/2\mp \nu}e^{-\pi\,\text{Im}\,\nu}\int_{\text{T}\{zz'\}}\Big[z^{-1\pm \nu}J_{342}^{++}(z')I_+(z)\text{H}_{\nu^*}^{(2)}(z) + z'^{-1\pm \nu}J_{342}^{+-}(z)I_+(z')\text{H}_{\nu}^{(1)}(z')\Big]\n\\
 &~-\int_{zz'}z'^{-1\pm \nu}J_{342}^{+-}(z)I_-(z')\text{H}_{\nu}^{(1)}(z'),
\end{align}
while $\mathcal{I}_t^{\pm}$ and $\mathcal{I}_u^{\pm}$ can be got from $\mathcal{I}_s^{\pm}$, again, by the substitution $2\leftrightarrow 3$ and $2\leftrightarrow 4$, respectively.

\begin{figure}[tbph]
\centering
\includegraphics[width=0.8\textwidth]{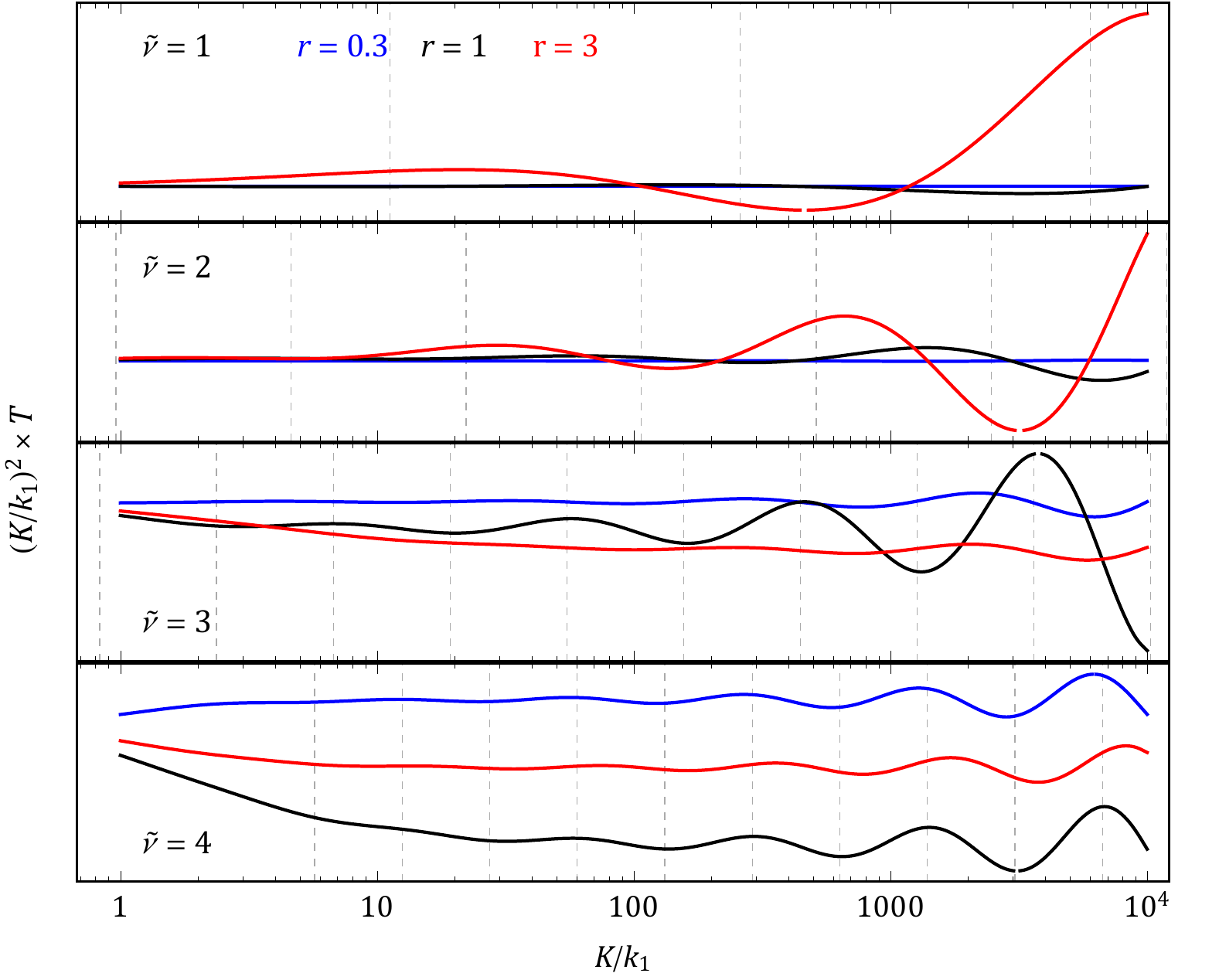}
\caption{The clock signals in the triangular limit of the trispectrum in QSFI with general sound speed, as function of momentum ratio $K/k_1$, with $k_1$ the soft momentum and $K=k_1+k_2+k_3+k_4$. The four rows correspond to four choices of $\wt\nu=\sqrt{(m/H)^2-9/4}$ and, in each row, the blue, black, and red curves show the signals for sound-speed ratio $r=c_\phi / c_\sigma = 0.3,1,3$, respectively.}
\label{Fig_TriangularClock}
\end{figure}

\begin{figure}[tbph]
\centering
\includegraphics[width=0.49\textwidth]{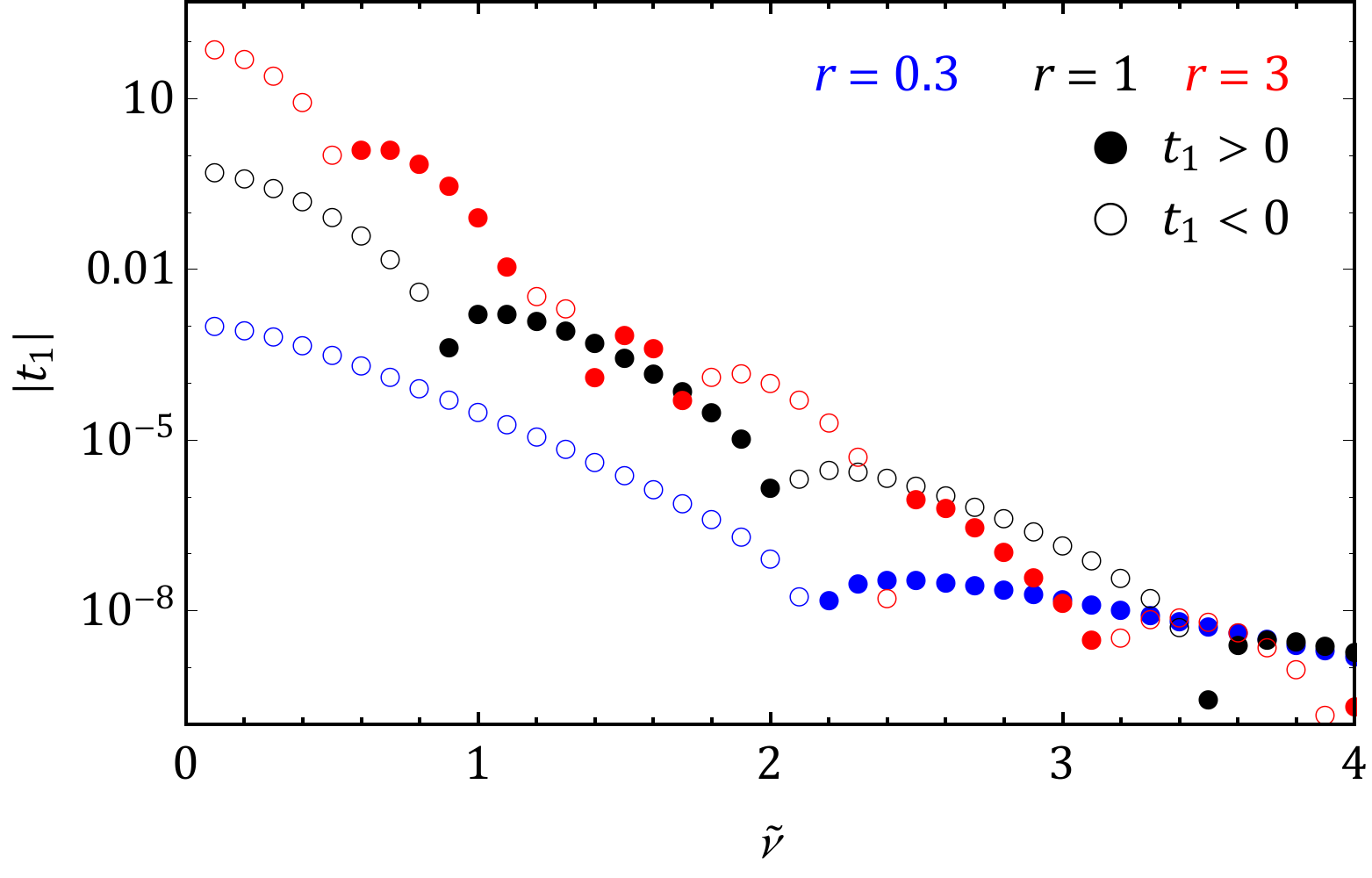}
\includegraphics[width=0.49\textwidth]{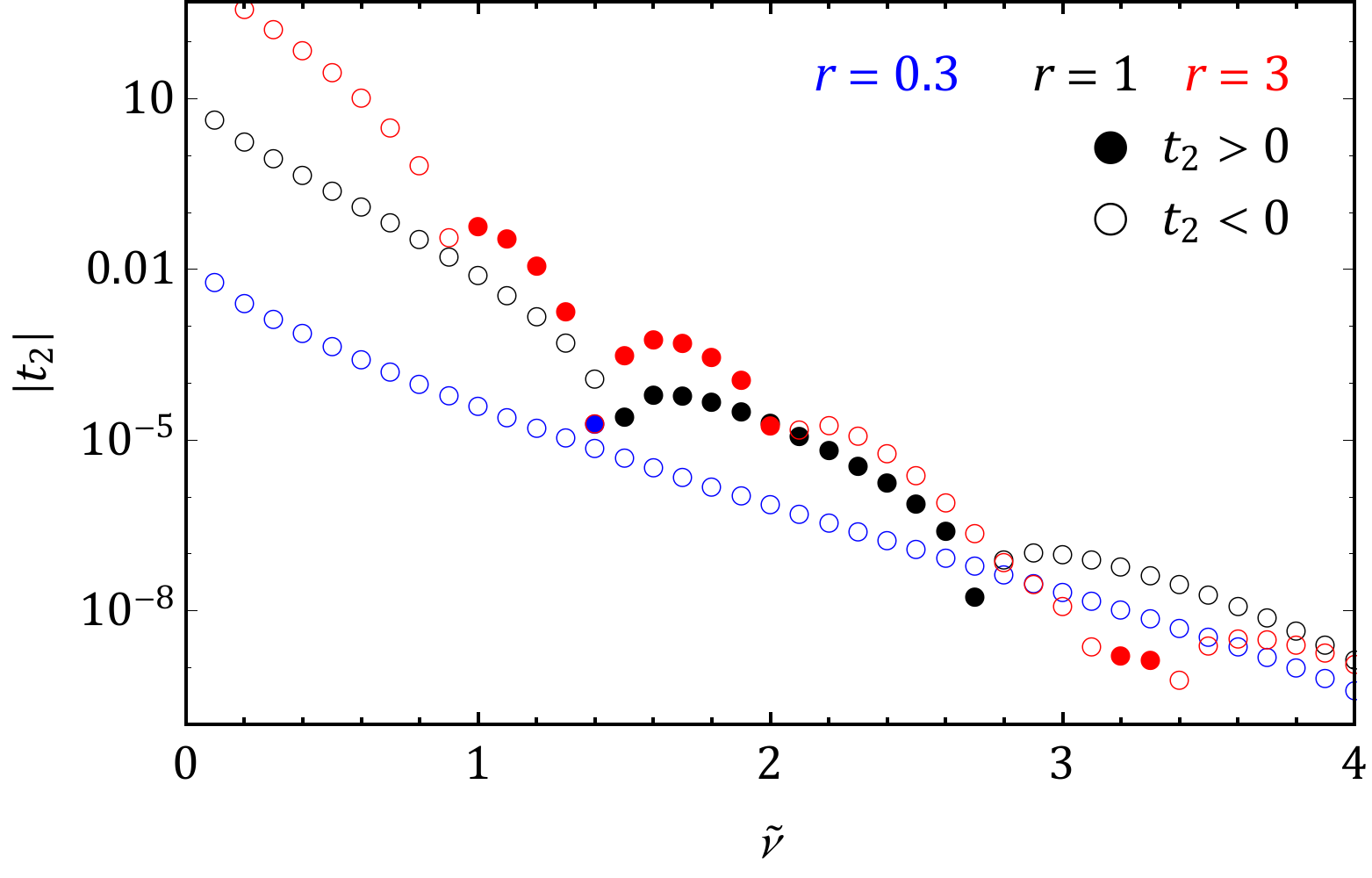}
\caption{The coefficient $t_1$ (left panel) and $t_2$ (right panel) for the oscillatory clock signal in the triangular limit of the trispectrum. The solid dots (empty circles) denote positive (negative) $t_{1,2}$. The blue, black, and red points correspond to sound-speed ratio $r=c_\phi / c_\sigma = 0.3,1,3$, respectively.}
\label{Fig_t1t2}
\end{figure}

In the case of purely imaginary $\nu$, we again have the oscillatory signals in the shape function in the $k_1/K\to 0$ limit, which can be expressed as,
\bge
\label{TriClock}
  T=\FR{\pi^3c_\phi^3 }{2^{15}P_\zeta} \Big(\frac{\ka_1}{H}\Big)^4\Big(\FR{k_1}{K}\Big)^{3/2}\Big[t_1\sin\Big(\wt\nu\log\FR{k_1}{K}\Big)+t_2\cos\Big(\wt\nu\log\FR{k_1}{K}\Big)+\cdots\Big],
\ede
where we have retained clock signals only, and $t_1=\text{Re}(t_+ + t_-)$ and $t_2=\text{Im}(t_- - t_+)$. We show in Fig.\;\ref{Fig_TriangularClock} the clock signals in the triangular limit of the trispectrum, for $\wt\nu=1,2,3,4$ and for $r=0.3,\,1,\,3$, respectively. The couplings are chosen to be $\lam_3=H$ and $\lam_4=\pi/4$ and the momentum configuration is chosen to be $k_1\ll k_2\simeq k_3\simeq k_4$, as shown in Fig.\;\ref{Fig_Triangular}. It is clear that the sound-speed ratio $r$ affects both the amplitude and the phase of the clock signal, while the frequency of the clock signal is independent of $r$, and is determined only by the mass $m$ of the massive mode $\de\si$. We further show in Fig.\;\ref{Fig_t1t2} the coefficients $t_{1,2}$ which control the amplitude of the clock signals. From Fig.\;\ref{Fig_t1t2} one can see clearly the Boltzmann suppression of the signal at large $\wt\nu$, while the degree of suppression is slightly different for different $r$.

\subsection{Collapsed Limit}

\begin{figure}[tbph]
\centering
\includegraphics[width=0.3\textwidth]{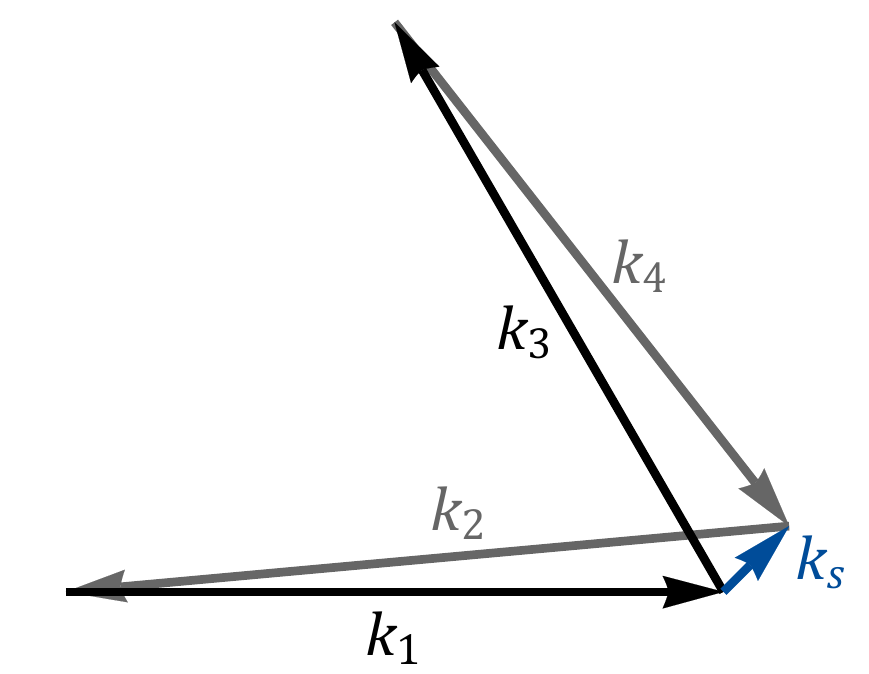}
\caption{The squeezed limit of the shape function.}
\label{Fig_Collapsed}
\end{figure}

\begin{figure}[tbph]
\centering
\includegraphics[width=0.8\textwidth]{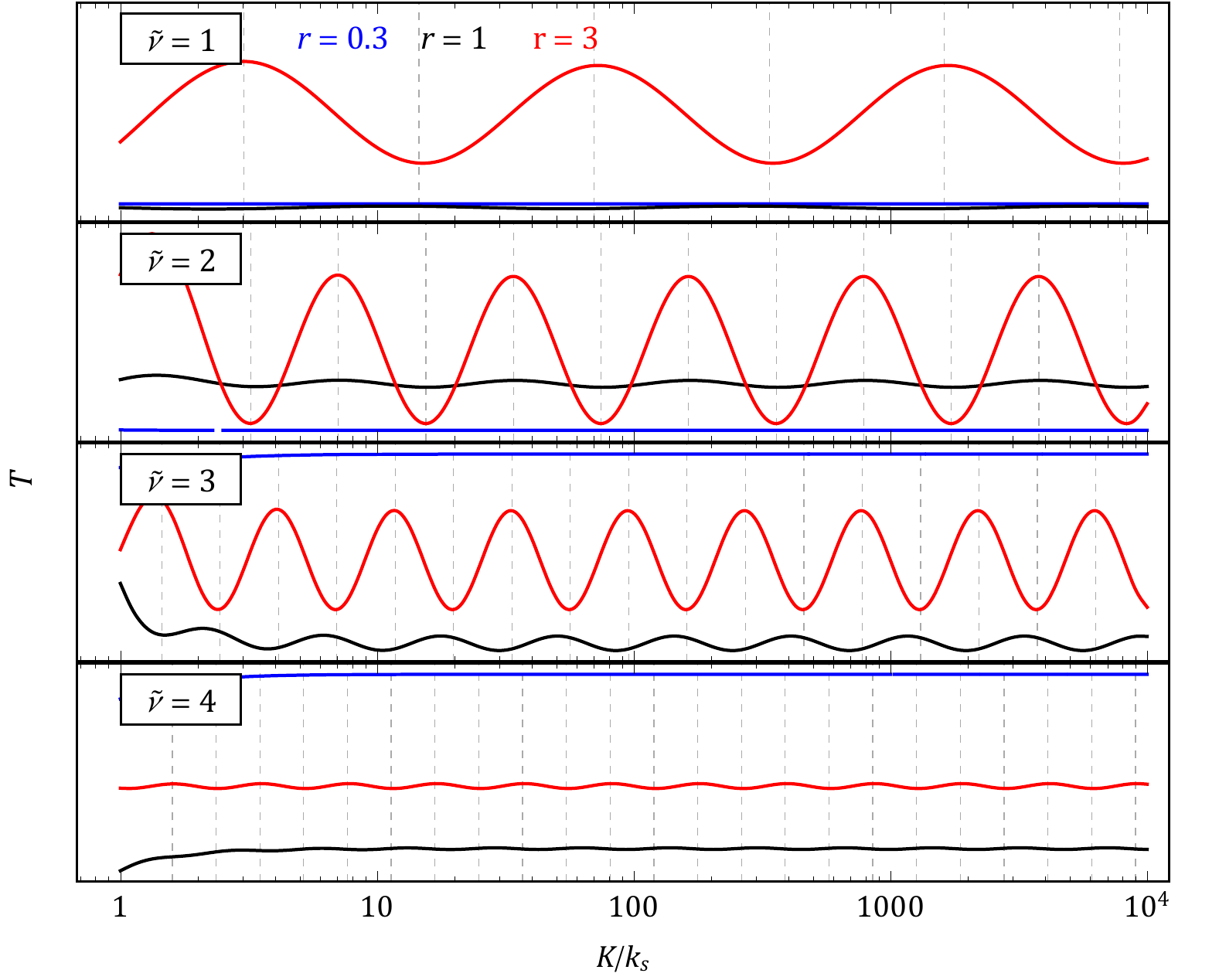}
\caption{The clock signals in the collapsed limit of the trispectrum in QSFI with general sound speed, as function of momentum ratio $K/k_s$, with $k_s$ the soft internal momentum and $K=k_1+k_2+k_3+k_4$. The four rows correspond to four choices of $\wt\nu=\sqrt{(m/H)^2-9/4}$ and, in each row, the blue, black, and red curves show the signals for sound-speed ratio $r=c_\phi / c_\sigma = 0.3,1,3$, respectively.}
\label{Fig_CollapsedClock}
\end{figure}

\begin{figure}[tbph]
\centering
\includegraphics[width=0.49\textwidth]{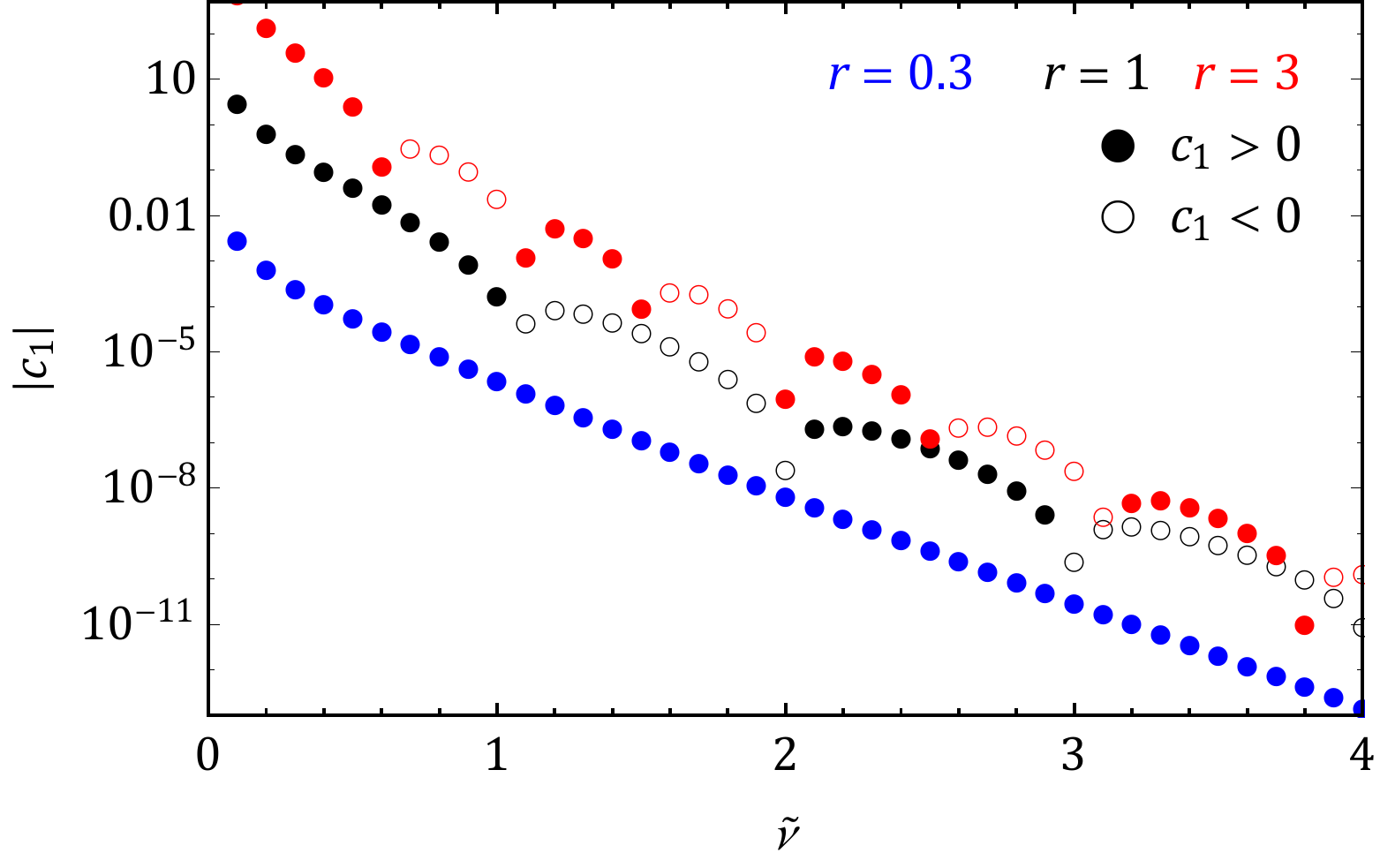}
\includegraphics[width=0.49\textwidth]{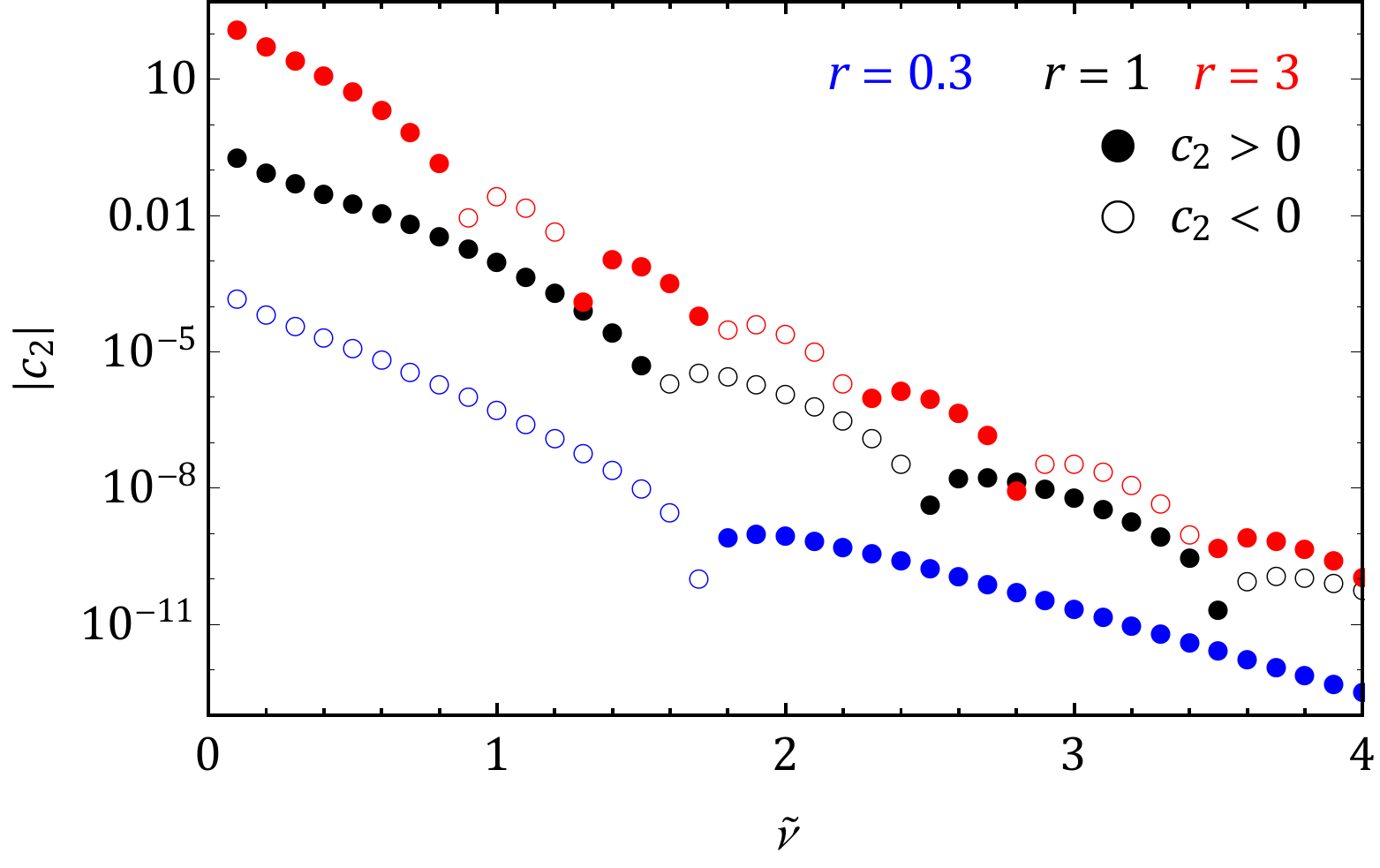}
\caption{The coefficient $c_1$ (left panel) and $c_2$ (right panel) for the oscillatory clock signal in the collapsed limit of the trispectrum. The solid dots (empty circles) denote positive (negative) $c_{1,2}$. The blue, black, and red points correspond to sound-speed ratio $r=c_\phi / c_\sigma = 0.3,1,3$, respectively.}
\label{Fig_c1c2}
\end{figure}

There is yet another possible soft limit for the 4-point function apart from the triangular limit, that is, the internal propagator in $\lam_3^2$-diagrams in (\ref{4ptdiag}) can also go soft. It is clear that only one of the three $\lam_3^2$-diagram can contribute a clock signal if all four external momenta $\mb k_i~(i=1,\cdots, 4)$ remain hard while one combination $|\mb k_i+\mb k_j|$ goes soft. We call such a configuration the collapsed limit and show an example of collapsed configuration in Fig.\;\ref{Fig_Collapsed}. Without loss of generality, we consider the case that $ k_s\equiv |\mb k_1+\mb k_2|$ becomes soft, where only the $s$-channel diagram contributes to the clock signal. To find this part, we need to expand the internal propagator of $\de\si$ field in the small momentum limit. From the expressions for the propagator (\ref{SKprop}) and the mode function of the massive field (\ref{modems}), we see that it amounts to expand a product of two Hankel functions as follows,
\begin{align}
  \text{H}_{\nu}^{(1)}(z)\text{H}_{\nu^*}^{(2)}(z')=\FR{e^{\pi\,\text{Im}\,\nu}}{\pi^2}\Big[\Gamma^2(\nu)\Big(\FR{zz'}{4}\Big)^{-\nu}+\Gamma^2(-\nu)\Big(\FR{zz'}{4}\Big)^{\nu}+\cdots\Big],
\end{align}
where we only keep terms proportional to $(zz')^{\pm\nu}$ which contribute to the clock signal, but not terms proportional to $(z/z')^{\pm\nu}$, which do not contribute to the clock signal. It's easy then to find that $(zz')^{\pm\nu}$ terms contribute in the trispectrum $T$ as,
\begin{align}
\label{TCollap}
&T=\FR{\pi^3c_\phi^3}{2^{15}P_\zeta}\Big(\frac{\ka_1}{H}\Big)^4\Big(\frac{\lambda_3}{H}\Big)^2\,\text{Re}\,\bigg[c_+\Big(\FR{k_s}{K}\Big)^{2\nu}+c_-\Big(\FR{k_s}{K}\Big)^{-2\nu}+\cdots\bigg],
\end{align}
with
\begin{align}
\label{cpm}
c_\pm
  =  &~ \FR{\Gamma^2(\mp\nu)r^5}{2^{5\mp \nu}\pi^2}\bigg(2+\FR{k_1}{k_3}+\FR{k_3}{k_1}\bigg)^{-5/2\pm\nu} \int_{zz'}(zz')^{-5/2\pm \nu}I_+^2( z )\Big[I_+^2 (z' )-I_-^2 ( z' )\Big],
\end{align}
where we have used $K\simeq 2(k_1+k_3)$ in the collapsed limit $k_s/K\ll 1$. Completely parallel to our previous examples in the squeeze limit of the bispectrum as well as the triangular limit of the trispectrum, we can extract the clock signal from (\ref{TCollap}), as,
\begin{align}
&T=\FR{\pi^3c_\phi^3}{2^{15}P_\zeta}\Big(\frac{\ka_1}{H}\Big)^4\Big(\frac{\lambda_3}{H}\Big)^2\Big[c_1\sin\Big(2\wt\nu\log\FR{k_s}{K}\Big)+c_2\cos\Big(2\wt\nu\log\FR{k_s}{K}\Big)+\cdots\Big],
\end{align}
where $c_1=\text{Re}(c_+ + c_-)$ and $c_2=\text{Im}(c_- - c_+)$. We show in Fig.\;\ref{Fig_CollapsedClock} the clock signals in the collapsed limit of the trispectrum, for $\wt\nu=1,2,3,4$ and for $r=0.3,\,1,\,3$, respectively. The momentum configuration is chosen to be $k_1\simeq k_2\simeq k_3\simeq k_4\simeq k_t\simeq k_u/\sqrt{3}$ while $k_s/K\ll 1$, as shown in Fig.\;\ref{Fig_Collapsed}. Again, the sound-speed ratio $r$ affects both the amplitude and the phase of the clock signal, while the frequency of the clock signal depends only on the mass $m$, but is twice of the frequency in the triangular limit. We further show in Fig.\;\ref{Fig_c1c2} the coefficients $c_{1,2}$ which control the amplitude of the clock signals. Again the Boltzmann suppression is manifest but the degree of suppression is dependent on the sound-speed ratio $r$, as is clearly seen in Fig.\;\ref{Fig_c1c2}.

While the generation mechanism of the clock signal in the case of the triangular limit is straightforwardly similar to that of the bispectrum \cite{Chen:2015lza,Chen:2016qce}, the mechanism in the case of the collapsed limit is slightly more complicated. Firstly, to have a clock field, the soft momentum has to be carried by the massive mode, which suggests that the collapse-limit clock signal is absent in the contact diagram and solely contributed by the scalar-exchange diagram. Secondly, this massive field resonates with two pairs of massless curvature modes and generate two sets of clock signals. The final clock signal is a result of interference between these two clock signals and hence has a doubled frequency. Finally, we note that the integrals in the coefficients (\ref{cpm}) of the clock signal in the collapsed limit are almost the square of the corresponding integral in the coefficients (\ref{3ptreform}) of the clock signal in the squeezed bispectrum, which resembles a consistency relation. In particular, we note that an additional clock signal is generated in the collapsed limit if we further take the squeezed limit $k_1/k_3\ll 1$, as is clear from (\ref{cpm}). However we do not expect exact match between the square of (\ref{3ptreform}) and (\ref{cpm}) since the soft leg in the collapsed limit is the massive propagator of $\de\si$ while the soft leg in the squeezed bispectrum is a mixed propagator. This relation was also derived in the special case of $r=1$ and $m<3H/2$ in \cite{Assassi:2012zq}.

\section{Discussions}
\label{sec_Disc}

In this paper we have calculate explicitly the primordial bispectrum and trispectrum in a typical model of QSFI with general sound speed. The calculation was carried out in a covariant diagrammatic approach of the in-in formalism, and was greatly simplified with the aid of the mixed propagator introduced in \cite{Chen:2017ryl}. We paid special attention to various soft limits of the correlation functions where the results display oscillatory quantum clock signals when the massive scalar mode has a mass $m$ greater than $3H/2$, although all our results can be analytically continued to the $m<3H/2$ regime where the predictions of QSFI take the non-trivial scaling behavior.

The measurement of trispectrum is typically more challenging than that of bispectrum, in terms of both the required number of observed modes and the amound of data procession. But the trispectrum also contain much richer information about the dynamics of inflation and the particle physics at the energy scale of inflation. Therefore it is important to know the strength of the clock signals calculated in this paper. The calculation in  previous sections made it clear that the amplitude of the clock signal is suppressed by the Boltzmann factor $e^{-\pi m/H}$ relative to the overall non-Gaussianity. So we are mostly interested in the case with $m\gtrsim H$, where the clock signal would have roughly the same order with the overall non-Gaussianity. 

Without addressing the question of to what degree the clock signals calculated in this paper can actually be measured, here we briefly comment on what can be learnt from the clock signals if both of their  amplitude and the frequency in all soft limits are well measured. Clearly, the measurement of the frequency $\wt\nu$ (or $2\wt\nu$ in the case of collapsed limit of the trispectrum) would tell us the mass $m$ of the massive field $\de\si$. With $\wt\nu$ known, one can in principle find the sound-speed ratio $r$ from the amplitude of the collapsed-limit clock signal, by varying the momentum configuration, since the amplitude $c_{1,2}$, or equivalently $c_\pm$ in (\ref{cpm}), depends on all external momentum, and the dependence is controlled by $\wt\nu$ and $r$. In a similar way, one can fit the cubic and quartic self-couplings, $\lam_3$ and $\lam_4$, of the massive field $\de\phi$, by exploiting momentum-dependence in the triangular limit of the trispectrum. Finally, one can find the bilinear coupling $\ka_1$ between the massive field and the inflaton, as well as the sound speed of the inflaton $c_\phi$, by comparing the amplitudes of clock signals in different limits. Moreover, the phase of the clock signal can be considered as a further consistency check of the above parameters.  Admittedly, the discussion here is highly ideal and measuring any parameter listed above in reality is hampered by a number of experimental issues, but it can still serve as a good case for more thorough study of the trispectrum.

The calculation in this paper has been done in the weak coupling regime where the dimension-1 bilinear mixing $\ka_1$ between the massive mode and the inflaton is smaller than the mass $m$ of the massive mode. On the other hand, the parameter space of the QSFI described by (\ref{SQSFI}) also contain an interesting region where the massive mode and the inflaton is strongly coupled in the sense that $\ka_1\gg m$, while other couplings remain weak and the theory is again well defined. It would also be interesting to study the non-Gaussianity more systematically in this strong coupling regime. To do that, one may seek for a generalization of the mixed propagator, or apply partial effective field theory methods to integrate out the local contributions and only calculate the non-local clock signal perturbatively \cite{Iyer:2017qzw}.

\paragraph{Acknowledgement.} We thank Shiyun Lu for her collaboration at the initial stage of this work. XC is supported in part by the NSF grant PHY-1417421. WZC is supported by the Targeted Scholarship Scheme under the HKSAR Government Scholarship Fund and the Kerry Holdings Limited Scholarship. YW is supported in part by ECS Grant 26300316 and GRF Grant 16301917 from the Research Grants Council of Hong Kong. ZZX is supported in part by Center of Mathematical Sciences and Applications, Harvard University.

\end{document}